\documentclass[aps,prl,reprint,groupedaddress]{revtex4-1}

\usepackage{graphicx}	
\usepackage{amsmath}	
\usepackage{amssymb}	

\usepackage{color}

\begin{document}

\newcommand{\imag}{\textrm{i}}
\newcommand{\sv}[1]{{\textcolor{blue}{\sf{[SV: #1]}}}}
\newcommand{\msun}{\ensuremath{M_\odot}}
\newcommand{\chieff}{\ensuremath{\chi_{\rm eff}}}
\newcommand{\chip}{\ensuremath{\chi_{\rm p}}}

\title{Prospects for Detecting Gravitational Waves at 5\,Hz with Ground-Based Detectors}

\author{Hang Yu}
\email{hyu45@mit.edu}
\affiliation{LIGO, Massachusetts Institute of Technology, Cambridge, Massachusetts 02139, USA}

\author{Denis Martynov}
\email{denism@mit.edu}
\affiliation{LIGO, Massachusetts Institute of Technology, Cambridge, Massachusetts 02139, USA}

\author{Salvatore Vitale}
\affiliation{LIGO, Massachusetts Institute of Technology, Cambridge, Massachusetts 02139, USA}

\author{Matthew Evans}
\affiliation{LIGO, Massachusetts Institute of Technology, Cambridge, Massachusetts 02139, USA}

\author{Bryan Barr}
\affiliation{SUPA, University of Glasgow, Glasgow G12 8QQ, United Kingdom}

\author{Ludovico Carbone}
\affiliation{School of Physics and Astronomy and Institute of Gravitational Wave Astronomy, University of Birmingham, Edgbaston, Birmingham B15 2TT, United Kingdom}

\author{Katherine L. Dooley}
\affiliation{The University of Mississippi, University, Mississippi 38677, USA}

\author{Andreas Freise}
\affiliation{School of Physics and Astronomy and Institute of Gravitational Wave Astronomy, University of Birmingham, Edgbaston, Birmingham B15 2TT, United Kingdom}

\author{Paul Fulda}
\affiliation{NASA Goddard Space Flight Center, Greenbelt, Maryland 20771, USA}

\author{Hartmut Grote}
\affiliation{Max-Planck Institut f\"ur Gravitationsphysik und Leibniz Universit\"at Hannover,
D-30167 Hannover, Germany}

\author{Giles Hammond}
\affiliation{SUPA, University of Glasgow, Glasgow G12 8QQ, United Kingdom}

\author{Stefan Hild}
\affiliation{SUPA, University of Glasgow, Glasgow G12 8QQ, United Kingdom}

\author{James Hough}
\affiliation{SUPA, University of Glasgow, Glasgow G12 8QQ, United Kingdom}

\author{Sabina Huttner}
\affiliation{SUPA, University of Glasgow, Glasgow G12 8QQ, United Kingdom}

\author{Conor Mow-Lowry}
\affiliation{School of Physics and Astronomy and Institute of Gravitational Wave Astronomy, University of Birmingham, Edgbaston, Birmingham B15 2TT, United Kingdom}

\author{Sheila Rowan}
\affiliation{SUPA, University of Glasgow, Glasgow G12 8QQ, United Kingdom}

\author{David Shoemaker}
\affiliation{LIGO, Massachusetts Institute of Technology, Cambridge, Massachusetts 02139, USA}

\author{Daniel Sigg}
\affiliation{LIGO Hanford Observatory, Richland, Washington 99352, USA}

\author{Borja Sorazu}
\affiliation{SUPA, University of Glasgow, Glasgow G12 8QQ, United Kingdom}

\date{Received 18 December 2017; revised manuscript received 9 February 2018; published 6 April 2018}

\begin{abstract}
We propose an upgrade to Advanced LIGO (aLIGO), named LIGO-LF, that focuses on improving the sensitivity in the 5-30\,Hz low-frequency band, and we explore the upgrade's astrophysical applications. We present a comprehensive study of the detector's technical noises and show that with technologies currently
under development, such as interferometrically sensed seismometers and balanced-homodyne readout, LIGO-LF can reach the fundamental limits set by quantum and thermal noises down to 5\,Hz. These technologies are also directly applicable to the future generation of detectors. 
We go on to consider this upgrade's implications for the astrophysical output of an aLIGO-like detector.
A single LIGO-LF can detect mergers of stellar-mass black holes (BHs) out to a redshift of $z\simeq6$ and would be sensitive to intermediate-mass black holes up to $2000\,M_\odot$. The detection rate of merging BHs will increase by a factor of 18 compared to aLIGO. Additionally, for a given source the chirp mass and total mass can be constrained 2 times better than aLIGO and the effective spin 3-5 times better than aLIGO.  Furthermore, LIGO-LF enables the localization of coalescing binary neutron stars with an uncertainty solid angle 10 times smaller than that of aLIGO at 30\,Hz, and 4 times smaller when the entire signal is used. 
LIGO-LF also significantly enhances the probability of detecting other astrophysical phenomena including the tidal excitation of neutron star $r$-modes and the gravitational memory effects. 
\end{abstract}

\pacs{}

\maketitle

\emph{Introduction.--}
The detection of gravitational waves (GWs) from coalescing binary black holes (BHs)~\cite{LSC:16, GW151226, LSC:17a, GW170608, LSC:17b} by Advanced LIGO (aLIGO)~\cite{aLIGO:15} and Advanced Virgo (aVirgo)~\cite{aVirgo:15} heralded the era of GW astrophysics. 
However, detecting binaries that are more massive and further away than the current BH catalog is challenging. Since the merger frequency decreases as the total mass of the binary increases, systems more massive than a few $\times100\,M_\odot$ will no longer lie in the most sensitive band of aLIGO. Intermediate-mass black holes (IMBHs) are an example of systems likely to be missed by aLIGO~\cite{ColemanMiller:04, Mandel:08, Graff:15, Veitch:15, Haster:16, 2017PhRvD..96b2001A} . 
At the same time, a pair of $30\,M_\odot$ BHs at $z=2$ will appear to have a total mass of $180\,M_\odot$ due to the cosmological redshift~\cite{Cutler:94}, illustrating the difficulties of detecting even the stellar-mass BHs at cosmological distances. Therefore, improving the low-frequency sensitivity plays a crucial role in extending both the mass and the spatial range of detectability. 

Another scientific goal of GW detectors is to enable multimessenger astronomy,
 as demonstrated by the detection of a merging neutron star (NS) binary in GW and the follow-ups by electromagnetic telescopes~\cite{GW170817, GW170817b}.
To help the subsequent observations, GW observatories need to provide the source location not only accurately but also quickly.
Since the time to merger scales with frequency $f$ as $f^{-8/3}$,
 if the error area can shrink small enough at a lower frequency,
the  location information can be sent out at a much earlier time.
Consequently, improving the low-frequency sensitivity allows for more timely follow-up observations. 

In this Letter we propose an upgrade to aLIGO (and its evolution A+~\cite{Lazzarini:16}) that enables a significant enhancement in sensitivity in the 5-30\,Hz band while maintaining high frequency performance.
This new design, dubbed ``LIGO-LF'', can be implemented on a timescale of $\sim\!\!10$ yr and serve as
 a pathfinder for later upgrades like the Voyager~\cite{Adhikari:17}
 and  next-generation detectors like the Einstein Telescope~\cite{Hild:10, Sathyaprakash:12} and Cosmic Explorer~\cite{Evans:17}.

\begin{figure}
 \includegraphics[width=0.48\textwidth]{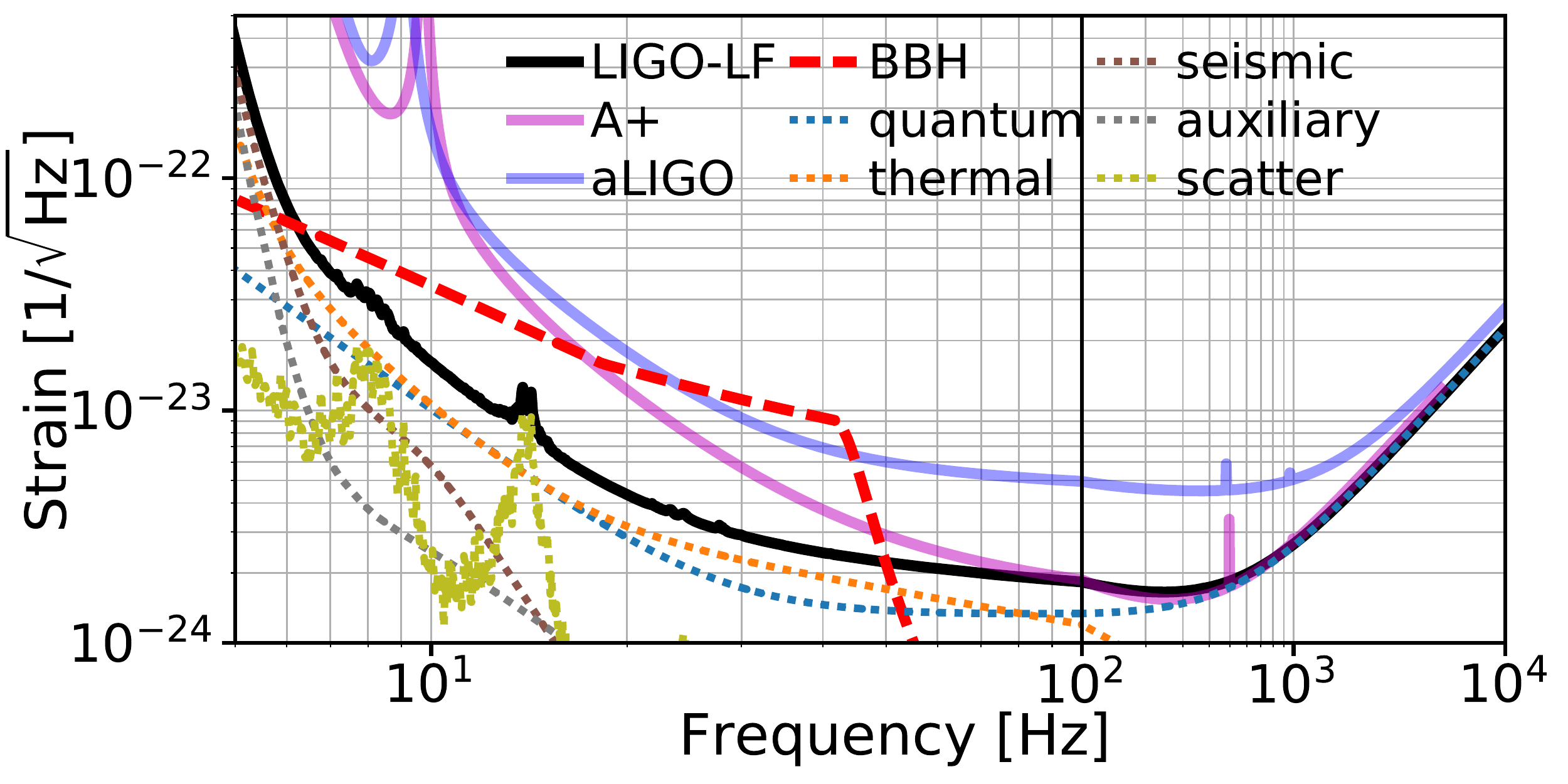}%
 \caption{Proposed sensitivity for LIGO-LF (solid black line) and its noise budget (dashed lines). Also shown in the dotted red curve is  the spectrum of a  $200\,M_\odot$-$200\,M_\odot$ binary BH merger (in the detector frame) at 10\,Gpc. LIGO-LF's sensitivity to such systems is greatly enhanced relative to aLIGO (solid blue line) and A+ (solid magenta line). Throughout this Letter, we will adopt the same coloring convention  when we compare different sensitivities (i.e., we use black, magenta, and blue for LIGO-LF, A+, and aLIGO, respectively).\label{fig:NB}}
\end{figure}

\emph{LIGO-LF design. --}
The current aLIGO sensitivity below 30\,Hz is limited by nonstationary technical noises~\cite{Martynov:15, Martynov:16, Hall:17}. Here we describe the solutions that we propose to reach the LIGO-LF sensitivity shown in Fig. \ref{fig:NB}. 

The first element of the upgrade reduces the angular control noise.
Angular motion of the optics is actively stabilized using wavefront sensors with a typical sensitivity of $5\times10^{-15} {\rm \,rad/\sqrt{Hz}}$~\cite{Barsotti:10, Martynov:16}. 
The bandwidths of the arm cavity angular loops are set to 3\,Hz to reduce the seismically induced motion to a few nrad rms. However, the control noise disturbs the test masses above 5\,Hz and contaminates the GW readout via beam miscentering on the mirrors. We propose to further suppress the motion of the optical benches so that the control bandwidth can be lowered. 

\begin{figure}
 \includegraphics[width=0.48\textwidth]{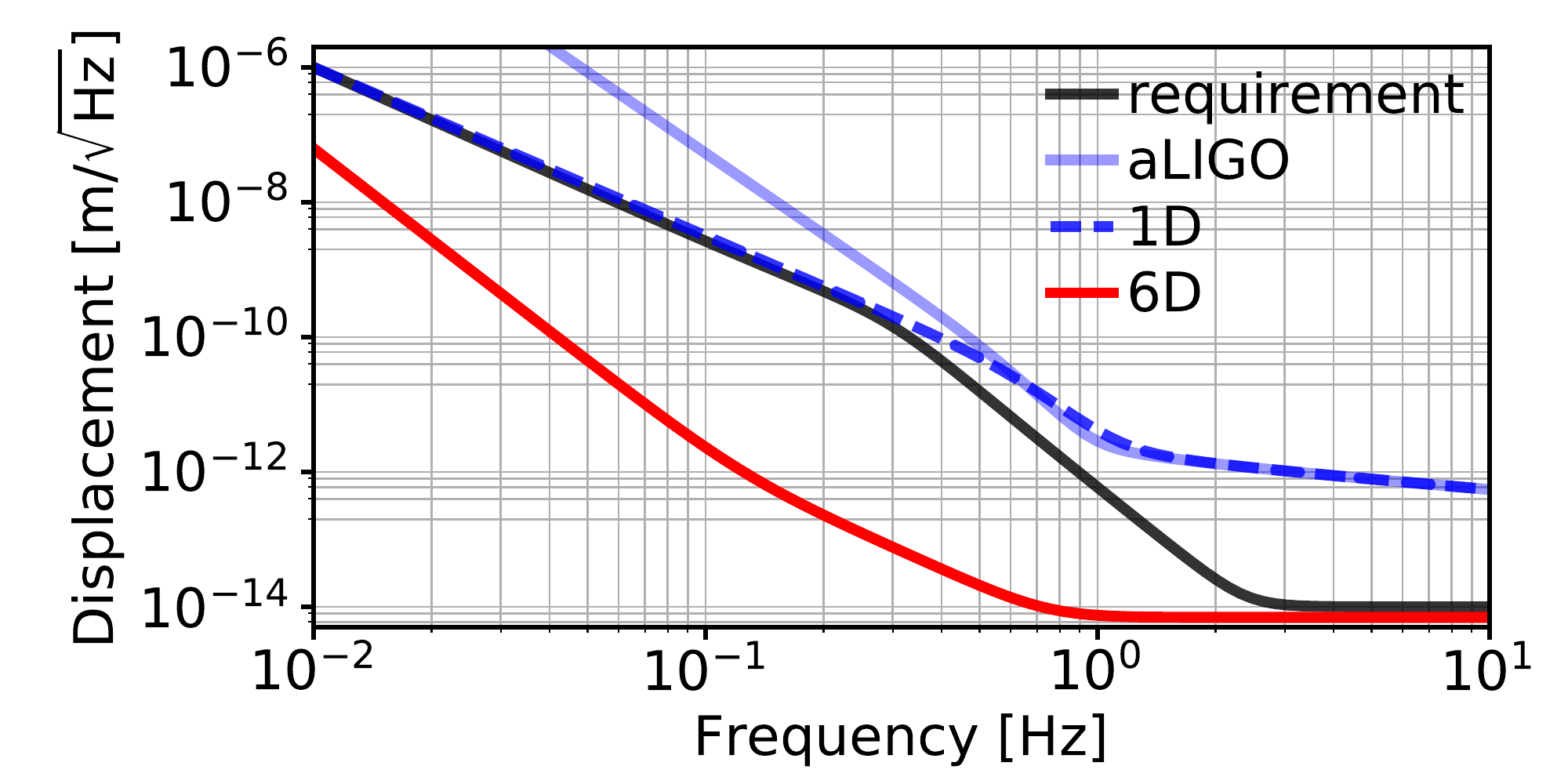}%
 \caption{Inertial sensor noise for aLIGO (blue line) and the requirement for LIGO-LF (black line). Custom tiltmeters can be used to improve aLIGO sensor noise below 0.5\,Hz (blue dashed line). A novel 6D seismometer (red line) can surpass the requirement in the entire band.
\label{fig:resi_gnd}}
\end{figure}

Despite the sophistication of LIGO's seismic isolation~\cite{Matichard:15, Matichard:15b, Matichard:15c}, it does not significantly reduce the microseismic motion at $\sim\!\! 0.2$\,Hz. This is due to tilt-to-horizontal coupling~\cite{Lantz:09, Matichard:15d, Matichard:16},
 which causes the noise of the aLIGO inertial sensors to grow as $1/f^{4}$ at low frequencies as shown in Fig.~\ref{fig:resi_gnd}.
To reduce the bandwidth of the angular controls to 1\,Hz, the tilt motion needs to be suppressed to $10^{-10} {\rm \,rad/\sqrt{Hz}}$ in the 0.01-0.5\,Hz band. The corresponding horizontal sensitivity is shown in Fig.~\ref{fig:resi_gnd}. Above 1\,Hz we require an improved sensitivity to reduce the direct coupling of the ground motion (see Supplemental Material~\cite{Supp} for a breakdown of the noise, which includes Ref~\cite{Hirose:10}).
 
There are two approaches to reach the required sensitivity of the inertial seismic sensors. The first one is to actively stabilize the tilt motion using custom-built tiltmeters~\cite{Venkateswara:14, Venkateswara:17}, which can achieve the requirement below 0.5\,Hz.
The second approach uses a novel 6D seismometer~\cite{Mow-Lowry:18}. In the core of this instrument is a quasimonolithically suspended~\cite{Cumming:12} mass whose position is monitored using an interferometric readout. Figure~\ref{fig:resi_gnd} shows that the design performance of the 6D seismometer satisfies the requirement in the entire band.

The radiation-pressure-induced angular instability also limits the minimum bandwidth~\cite{Sidles:06, Dooley:13}. We propose to increase LIGO-LF's test masses from 40 to 200\,kg to mitigate the effects of radiation pressure. 
More massive test masses are also a fundamental part of next-generation GW detectors.

The coupling of the longitudinal motion of the signal recycling cavity contaminates aLIGO's sensitivity in the 10-50\,Hz band~\cite{Martynov:16}. This coupling is proportional to the arm detuning~\cite{Izumi:17} introduced to enable the dc readout of the GW signal~\cite{Fricke:12}. For LIGO-LF, we assume balanced-homodyne readout~\cite{Fritschel:14} will be implemented instead, which essentially eliminates the coupling.

In aLIGO, high-quality-factor suspension resonances are damped using shadow sensors~\cite{Carbone:12} with a sensitivity of $2\times10^{-10} {\rm \,m/\sqrt{Hz}}$.
A global control scheme has been proposed~\cite{Shapiro:15} to reduce its direct coupling to the GW output.
However, this noise still enters the auxiliary  loops and couples to the GW output indirectly.
This calls for an improvement of the sensor noise by a factor of 100.
Interferometric sensors~\cite{Aston:11} are promising candidates
 and are used in the LIGO-LF design.

Once technical noises are suppressed, LIGO-LF sensitivity will be limited by quantum and thermal noises. Our strategy to improve the fundamental limits is similar to the Strawman Team Red design~\cite{Hild:12}.  

Quantum noise~\cite{Buonanno:01, Miao:12, Martynov:17} manifests both as sensor shot noise and as displacement noise by exerting quantum radiation pressure (QRP) forces on the test masses. LIGO-LF will operate under ``resonant-sideband extraction''~\cite{Mizuno:93} with the same amount of power circulating in the arms as aLIGO. A signal recycling mirror transmissivity of $0.25$ is chosen to optimize the broadband sensitivity.

The quantum noise can be further reduced with squeezed light~\cite{McClelland:11, LSC:11, LSC:13}. Here we assume a frequency-dependent squeezing~\cite{Kimble:01, Harms:03, Kwee:14, Oelker:2016bg}
that provides 3\,dB reduction of the QRP and 6\,dB of the shot noise.
Relative to aLIGO, QRP is further suppressed by the heavier test masses mentioned above.

Thermal noise~\cite{Saulson:90} from the suspension~\cite{Gonzalez:00, Cumming:12} and the optical coatings~\cite{Levin:98, Hong:13, Yam:15, Gras:17, Martynov:17} dominates the sensitivity from 5 to 100\,Hz. Suspension thermal noise can be lowered by doubling the length of the last suspension stage to 1.2\,m~\cite{Young:02, Hammond:12} and by applying more sophisticated surface treatments~\cite{Mitrofanov:03}. 
 LIGO-LF's penultimate masses will also need to be suspended with fused silica fibers to avoid excess noise.
 Furthermore, the vertical suspension resonance can be shifted down to 4.3\,Hz by increasing the fiber tension to 1.7\,GPa.
Overall, 
 a factor of 5 improvement over aLIGO suspension thermal noise is possible 
 (details of the LIGO-LF suspension are available in Supplemental Material~\cite{Supp}, including Refs~\cite{Aston:12, Rakhmanov:00, Fritschel:02}).

The larger test masses and better seismic isolation open up the possibility of increasing spot sizes on the test masses by $50\%$, with a corresponding reduction in the coating thermal noise.
Furthermore, a factor of 2 improvement in the coating loss angle is expected by the time of LIGO-LF~\cite{Steinlechner:16}. 

Further sensitivity improvement below 30\,Hz is limited by gravity gradient noise~\cite{Saulson:84, Hughes:98, Driggers:12, Creighton:08, Harms:15}.
It can be mitigated with offline regression~\cite{Coughlin:16}, and in our calculation we assume a factor of 10 cancellation~\cite{Evans:17}.
The residual is combined with the residual seismic motion in Fig.~\ref{fig:NB} under the label ``seismic''.

Scattering is another critical noise source below 30\,Hz~\cite{Flanagan:94, Ottaway:12, Martynov:15}. A small amount of light can scatter off the test masses due to surface imperfections, hit the baffles along the beam tubes, and finally recombine with the main beam. 
These stray photons induce differential power fluctuations which perturb the test masses via radiation pressure. 
In Fig.~\ref{fig:NB}, we present a scattering noise curve estimated from the typical ground motion at the LIGO sites with an anticipated 50\% improvement in the mirror surface quality relative to aLIGO. As the relative displacement between the test mass and the tube is comparable to the laser wavelength (1\,$\mu$m), the coupling can become nonlinear,  up-converting the baffle motion below 0.4\,Hz up to 5\,Hz~\cite{Canuel:13,Martynov:16} .
For rare cases where the ground motion is severe, an up-conversion shelf can form~\cite{Martynov:15} and limit the low-frequency sensitivity. The antireflection surfaces along the optical path also create scattering noise. To suppress it, baffles should be constructed to block $99.9\%$ of the stray light (details available in Supplemental Materials~\cite{Supp} with Ref~\cite{llo:29665}).

In summary, the key LIGO-LF advancements consist of low-noise, interferometric sensors for seismic isolation and suspension damping, and heavy test masses with large spot sizes for improving the fundamental limits. The LIGO-LF suspension system is also redesigned. Combined with the squeezed light, balanced-homodyne readout, and low-loss coating that are planned for A+, the upgrades lead to the final LIGO-LF sensitivity.

\emph{Astrophysical applications.--} 
LIGO-LF can deliver a rich array of science in astrophysics. Here we consider three examples: (i) binary BHs, including the expected range of detectability and detection rate, and parameter estimation (PE) of events, (ii) binary NSs, focusing on the source localization and the detectability of the tidal excitation of NS $r$-modes, and (iii) the GW memory effect. The technical details are provided in Supplemental Material~\cite{Supp} with Refs~\cite{Planck:16, 2016ApJ...818L..22A, Jaranowski:98, Vitale:17, Lai:94, Reisenegger:94, Douchin:01, Berti:06}. The searches for the stochastic GW background~\cite{Allen:97} and the continuous GW~\cite{Bejger:17} rely mostly on the instrument's high-frequency performance, and are not enhanced by LIGO-LF.

\begin{figure}
 \includegraphics[width=0.48\textwidth]{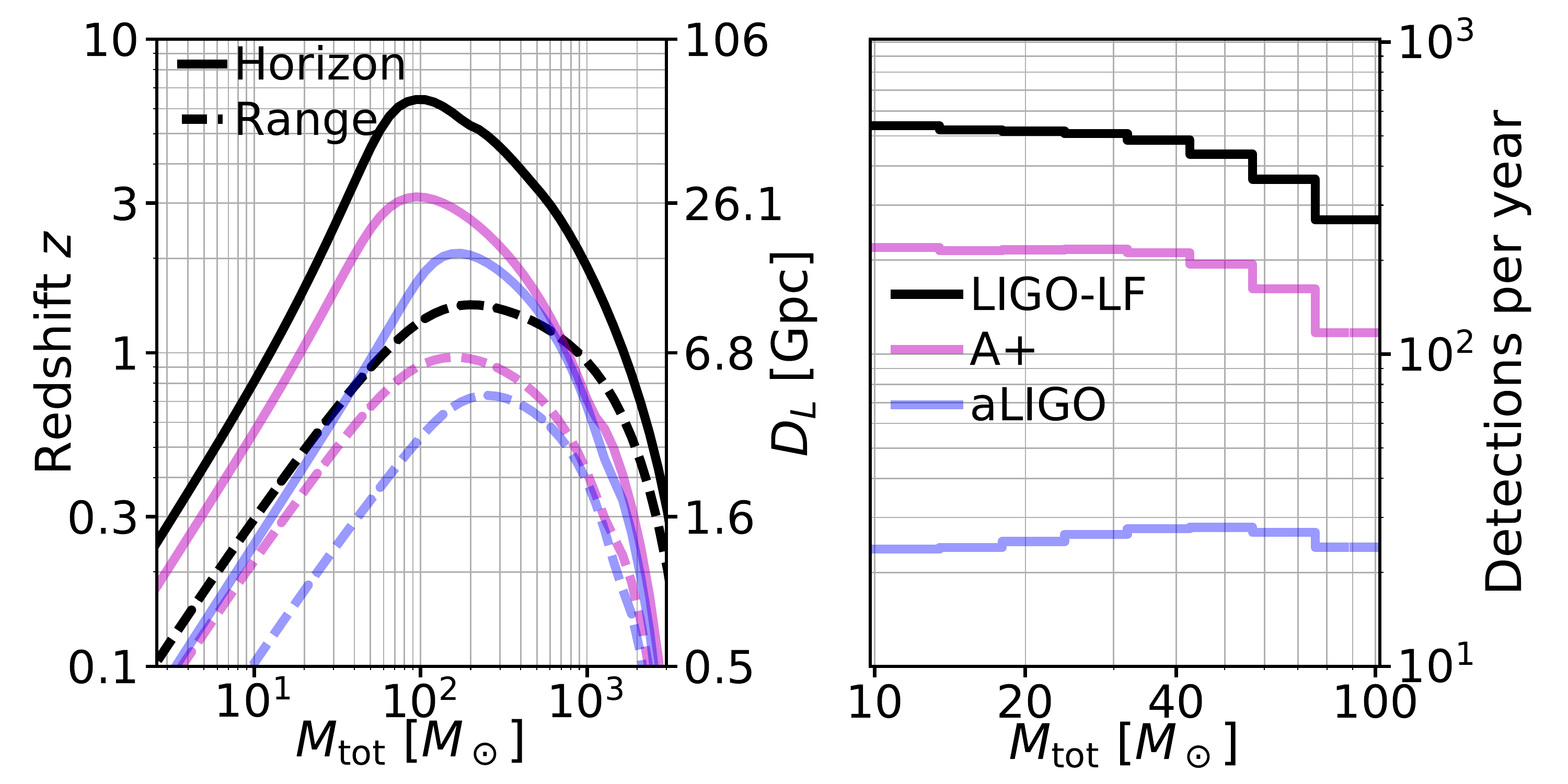}%
 \caption{Left: The horizon (solid line) and range (dashed line) for binaries with different (source-frame) total masses. A single LIGO-LF may reach a cosmological redshift of $z\simeq 6$. Right: Expected detections rate of coalescing stellar-mass BH binaries as a function of the total mass. We divide $M_1$ and $M_2$ each into eight logarithmic bins from $10\,M_\odot$ to $100\,M_\odot$ and marginalize over the mass ratio to derive the event rate per total mass bin. LIGO-LF can detect $\sim 4000$ events per year, 18 times more than the expected number for aLIGO.  All the numbers are calculated assuming a single detector.\label{fig:range}}
\end{figure}

\emph{(i).--}With the LIGO-LF upgrade, both the maximum detectable distance and mass and the number of detections are larger than with aLIGO and A+, as illustrated in Fig.~\ref{fig:range}. 
In the left, we plot the single-detector horizon and range~\cite{Chen:17} (in both redshift $z$ and luminosity distance $D_{ L}$) for binaries with different total masses.
The systems are assumed to be nonspinning and to have equal masses. A single LIGO-LF could detect binary BHs to cosmological distances ($z\simeq 6$), whereas a network of four detectors would observe to $z\sim10$, potentially accessing the first generation of stellar BHs~\cite{Sesana:09}. 

Assuming a power-law mass distribution and a merging rate of $97(1+z)^2\,{\rm Gpc^{-3}\,yr^{-1}}$~\cite{Ely:17, O1BBH:16}, the expected number of detections of coalescing BH binaries is shown in the right in Fig.~\ref{fig:range}. It predicts that a single LIGO-LF can detect $\sim4000$ merging BHs per year, 18 (2.3) times aLIGO's (A+'s) detection rate.  The large number of events observed by LIGO-LF increases the statistical signal-to-noise ratio (SNR), which may be used to separate formation channels that predict different event rates~\cite{Belczynski:10, Rodriguez:16b} and to constrain the fraction of dark matter in the form of primordial BHs~\cite{Bird:16, Ely:17}.

Moreover, LIGO-LF enables more accurate PE than aLIGO.
To emphasize the improved low-frequency sensitivity, we consider binaries with detector-frame total mass $M^{(d)}_{\rm tot}\ge100\,M_\odot$. Since the sensitivity of A+ and aLIGO is similar below 20\,Hz, we consider the comparison between LIGO-LF and aLIGO.  Qualitatively, the improvements are due to two facts: A more total SNR is accumulated in LIGO-LF than in aLIGO, and the SNR starts to accumulate at lower frequencies. Thus, if aLIGO can measure only the merger-ringdown phase of a coalescence, with LIGO-LF we could access the inspiral phase as well, allowing for a more precise estimation of the component masses and spins.

To quantify these improvements, we simulate GW signals with the \texttt{IMRphenomPv2} waveform~\cite{Hannam:14} and inject them to mock detector noise.   
We consider five total mass bins from $100\,$\msun{} to $2000\,$\msun, each with three spin configurations: ($\chi_{\rm eff}$=$\chi_{p}$=$0$), ($\chi_{\rm eff}$=$0.5$, $\chi_{ p}$=$0.6$), and ($\chi_{\rm eff}$=$-0.5$, $\chi_{\rm p}$=$0.6$). Here \chieff{} is the mass-weighted sum of component spins along the orbital angular momentum~\cite{Ajith:11,Santamaria:10}, and \chip{} captures the precessing components~\cite{Schmidt:15}. The effect of the mass ratio has been studied in Refs.~\cite{Veitch:15, Haster:16}, so we focus on the equal mass case. 
We consider a four-detector network formed by the Hanford (H) and the Livingston (L) sites, LIGO-India (I), and aVirgo (V). For HLI, we consider both the LIGO-LF and aLIGO sensitivities; for V, we fix it at its design sensitivity~\cite{aVirgo:15}. 
KAGRA~\cite{Somiya:12} is not included as it is less sensitive to IMBHs. For each source, the inclination is fixed to $30^\circ$ and the distance is chosen such that the network SNR is 16 with aLIGO's sensitivity. We then use the \texttt{LALInference}~\cite{Veitch:15b} to get posterior distributions of source parameters. 
The PE results refer to the detector frame and we denote them with a superscript `($d$)'.

\begin{figure}
\begin{minipage}{0.5\textwidth}
\includegraphics[width=1\textwidth]{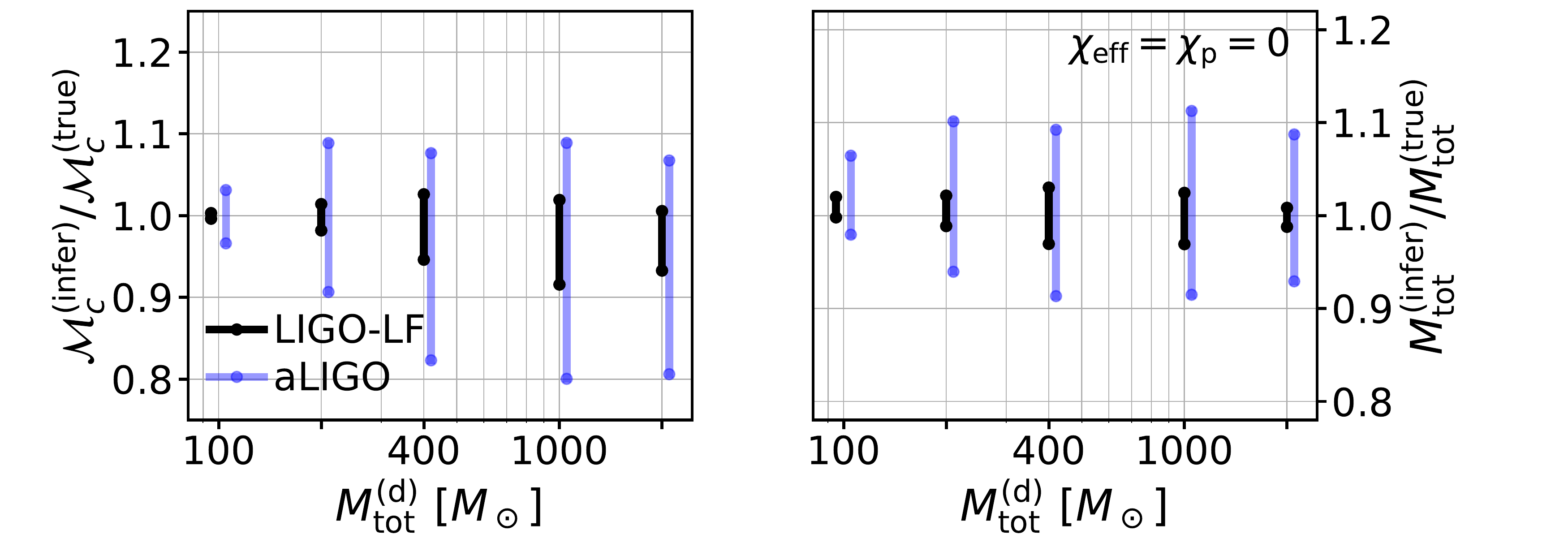}
\end{minipage}
\begin{minipage}{0.5\textwidth}
\includegraphics[width=1\textwidth]{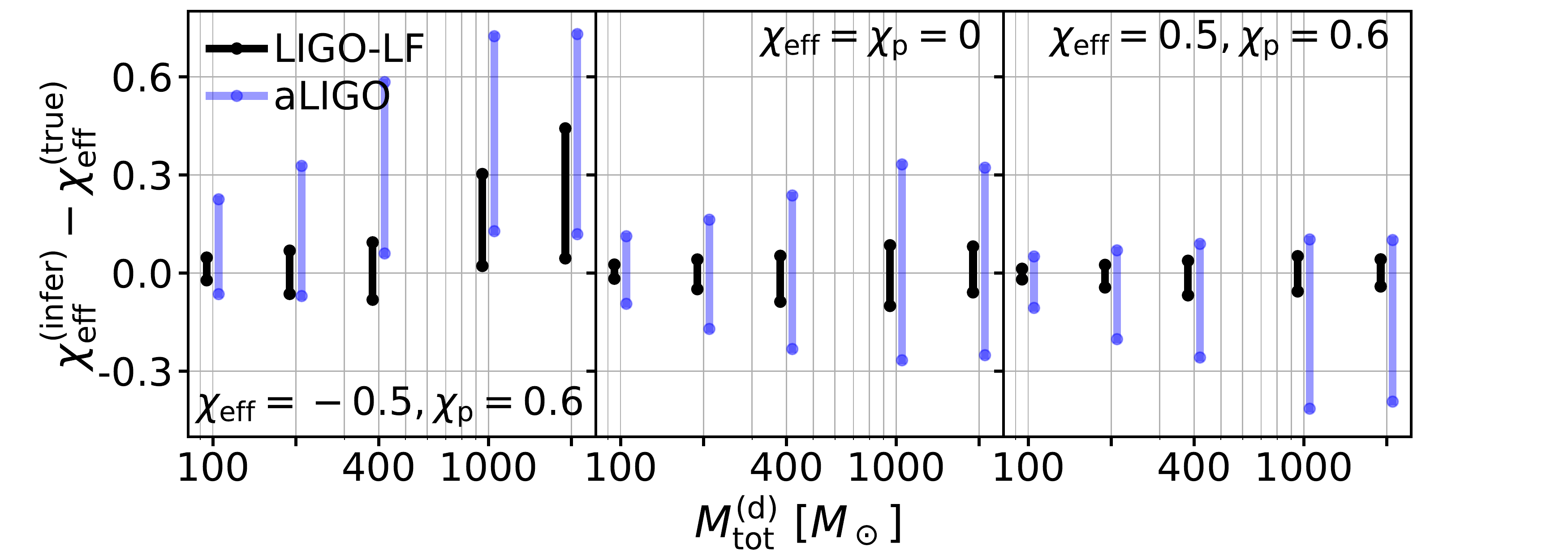}
\end{minipage}
\caption{The $90\%$ credible intervals of the detector-frame chirp mass $\mathcal{M}_c^{ (d)}$ (top left), total mass $M_{\rm tot}^{ (d)}$ (top right), and effective spin $\chi_{\rm eff}$ (bottom) are all significantly smaller for LIGO-LF than for aLIGO. LIGO-LF also reduces biases, especially for $\mathcal{M}_c^{ (d)}$ and $\chi_{\rm eff}$ when the spin is antialigned (bottom left).}
\label{fig:PE}
\end{figure}

In Fig.~\ref{fig:PE}, we plot the $90\%$ credible intervals of the chirp mass $\mathcal{M}_c^{(d)}$, total mass $M_{\rm tot}^{(d)}$, and $\chi_{\rm eff}$. For the masses, we present the results for the nonspinning case. When spins are included, an aligned (antialigned) spin tends to improve (degrade) the inference accuracy~\cite{Ng:18}. Similar effects can also be seen in the posterior distributions of $\chi_{\rm eff}$, as illustrated in the bottom panels. The precession term $\chi_{p}$ cannot be well constrained even with LIGO-LF. 

LIGO-LF typically enables a factor of $2$ improvement in constraining the sources' redshift compared to aLIGO, limiting the improvement in measuring the source-frame masses to a similar level (see Supplemental Material~\cite{Supp} for the redshift posteriors). 
The effective spin, nonetheless, is unaffected by the redshift and thus LIGO-LF can achieve 3-5 times better accuracy than aLIGO, which will be essential for discriminating between different formation scenarios that predict different spin configurations~\cite{Rodriguez:16a, Farr:18}. 

\begin{figure}
 \includegraphics[width=0.48\textwidth]{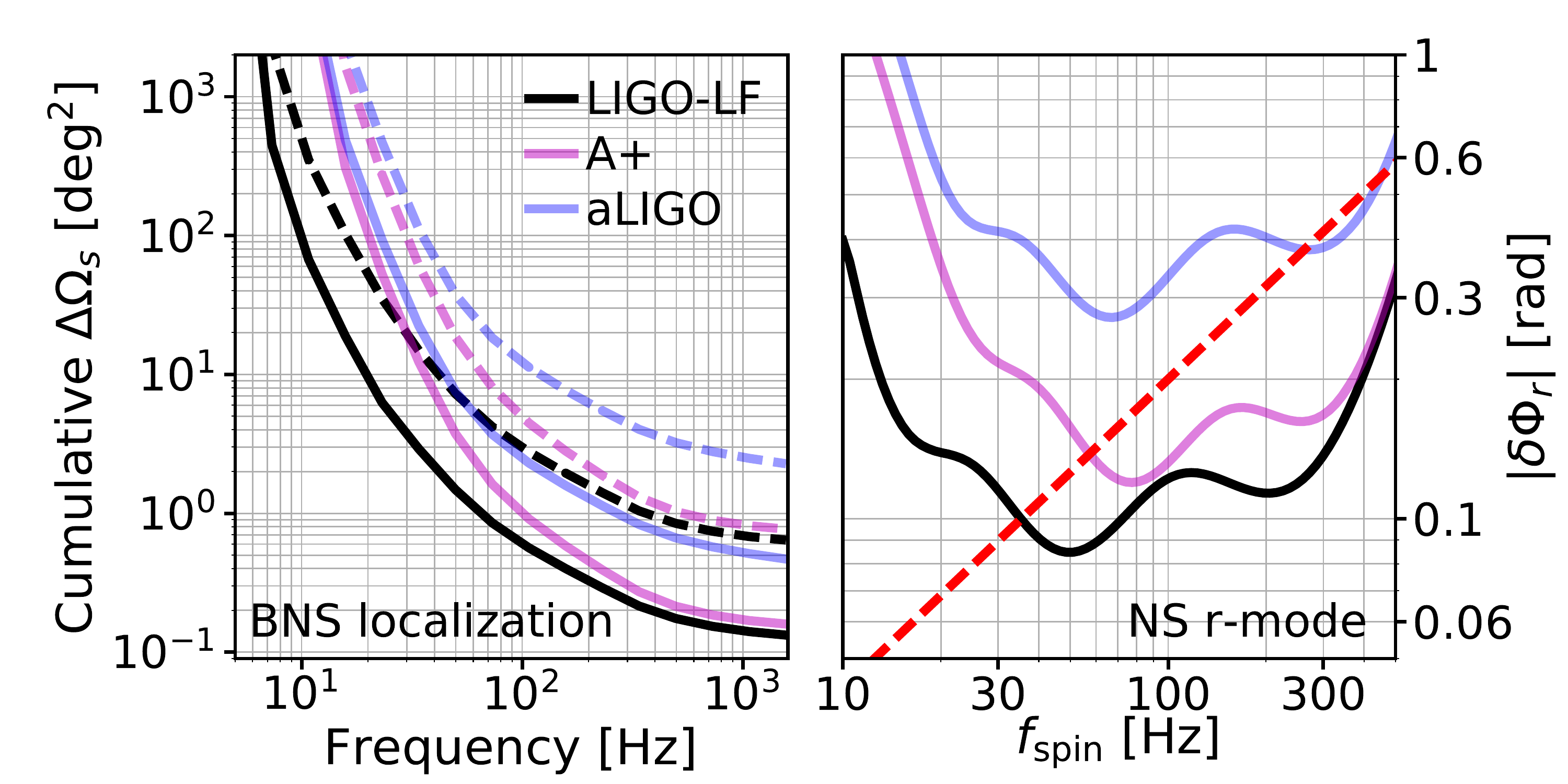}%
 \caption{Left: The cumulative uncertainty in localization, $\Delta \Omega_{s}$, for the HLIV network. We consider NS binaries at the Coma cluster with two inclinations, $30^\circ$ (solid line) and $75^\circ$ (dashed line),  and marginalize over the polarization and time of arrival. 
 LIGO-LF improves the localization by a factor of 3.5 over aLIGO using the entire signal and by a factor of 5 over A+ using only the sub-30\,Hz data. 
Right: The uncertainty (solid) in measuring the phase shift due to resonant excitation of the NS $r$-mode $\delta \Phi_r$ as a function of the NS spin frequency $f_{\rm spin}$. We consider the single detector case and fix the sources at 50\,Mpc with optimal orientation. 
Also shown in the red dashed line is the expected physical $r$-mode phase shift. The effect is detectable when the real phase shift is greater than the statistical error.
\label{fig:BNS}}
\end{figure}

\emph{(ii)--}We use the Fisher matrix to examine LIGO-LF's ability to localize a binary NS including effects of Earth's rotation~\cite{Wen:10, Zhao:17}. We consider the same network as in the PE section. The result is shown in the left panel in Fig. \ref{fig:BNS}. The final localization error in solid angle, $\Delta \Omega_{s}$, is 3.5 (1.2) times smaller with LIGO-LF than with aLIGO (A+). While LIGO-LF's improvement over A+ is mild when the entire signal is used, it is nonetheless dramatic (a factor of 5 over A+ and 10 over aLIGO) if we use only data below 30\,Hz, about 1\,min prior to the final coalescence. This illustrates LIGO-LF's ability to achieve a more timely localization than A+ and aLIGO. 

The $r$-mode study follows Ref.~\cite{Flanagan:07}, and we focus on the $l$=2, $m$=1 mode. The results are summarized in the right panel in Fig.~\ref{fig:BNS}. We find that if the NS spins at a rate greater than 35\,Hz~\cite{Burgay:03}, a single LIGO-LF may detect the $r$-mode resonance up to a distance of 50\,Mpc. Since the phase shift of the $m$=1 $r$-mode depends on the NS stratification, which is sensitive to the internal composition and the state of matter~\cite{Yu:17a, Yu:17b}, a detection may thus place constraints on the NS equation of state from physics beyond the star's bulk properties~\cite{Andersson:18}. Furthermore, the $r$-mode resonance provides an independent measurement of the NS spin, which may help break the spin-mass ratio degeneracy~\cite{Cutler:94} and improve the accuracy in measuring the (equilibrium) tidal deformability~\cite{Hinderer:10, GW170817}. 

\begin{figure}
 \includegraphics[width=0.48\textwidth]{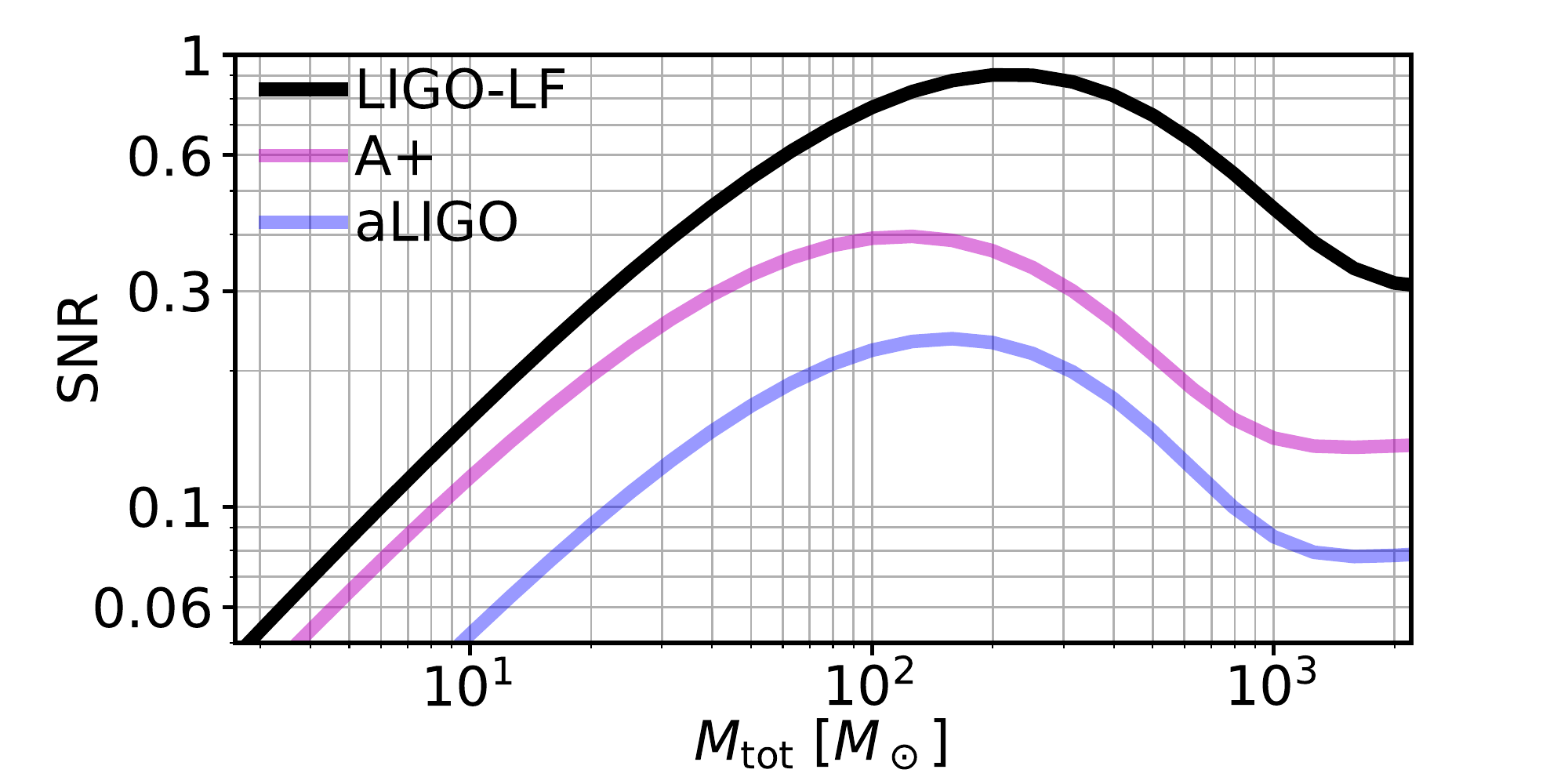}%
 \caption{SNR from the GW memory effect as a function of the source-frame total mass. The sources are fixed at $z=0.1$ and an inclination of $30^\circ$. The peak SNR seen in LIGO-LF is 4 (2) times greater than that seen in aLIGO (A+). 
\label{fig:GW_mem}}
\end{figure}

\emph{(iii)--}We consider the GW memory effect~\cite{Christodoulou:91} adopting the minimal-waveform model~\cite{Favata:09}. The result is shown in Fig.~\ref{fig:GW_mem}. Together with the increased detection rate, LIGO-LF has a promising probability to detect this effect via event stacking~\cite{Lasky:16}. 

\emph{Conclusions.--}
In this Letter, we propose LIGO-LF, an upgrade improving 
 aLIGO's low-frequency performance.
 The new technologies required for this update are directly applicable to the future generation of detectors.
Comparing LIGO-LF to aLIGO, the mass and spatial range of binary BHs detectable are greatly enhanced, and the localization of NS binaries can be achieved at a much earlier time, enabling more timely follow-up.

\begin{acknowledgments}
The authors thank Rainer Weiss, Peter Fritschel, Valery Frolov, Riccardo DeSalvo, Rana Adhikari, Christopher Berry, members of the LSC ISC and AIC groups, and the referees for the valuable discussions and comments. 
The authors acknowledge the support of the National Science Foundation (NSF) and the Kavli Foundation. LIGO was constructed by the California Institute of Technology and Massachusetts Institute of Technology with funding from the NSF and operates under Cooperative Agreement No. PHY-0757058. Advanced LIGO was built under Award No. PHY-0823459.
This material is based upon work supported by the NSF under Grant No. 1608922.  
H. Y. is supported in part by NASA ATP Grant No. NNX14AB40G (PI Nevin Weinberg). 
D. M. is supported by the Kavli Fellowship. 
L. C., C. M.-L., and A. F. have been supported by the Science and Technology Facilities Council (STFC).
B. B., G. H., S. Hild, J. H., S. Huttner, S.R., and B.S. are supported by the STFC (ST/N005422/1) and the International Max Planck Partnership (IMPP).
\end{acknowledgments}

\bibliography{refL}

\end{document}


\newcommand{\imag}{\textrm{i}}
\newcommand{\diff}{\textrm{d}}
\newcommand{\vect}[1]{\boldsymbol{#1}}


\title{Supplemental Material for ``Prospects for Detecting Gravitational Waves at 5\,Hz with Ground-Based Detectors''}


\author{}
\affiliation{}


\date{Received 18 December 2017; revised manuscript received 9 February 2018; published 6 April 2018}


\pacs{}

\maketitle

\section{LIGO-LF Suspension Design}\label{sec:sus}
LIGO-LF adopts a 4-stage suspension system similar to that of aLIGO~\cite{Aston:12}. The suspension chain consists of a top mass (TOP), an upper-intermediate mass (UIM), a penultimate mass (PUM), and a main test mass (TST), with the parameters for each stage summarized in TABLE \ref{tab:sus}. The blade design used for LIGO-LF vertical support is similar to that of aLIGO. Two requirements are set for the system above 5\,Hz: the suspension needs to provide sufficient filtering of the residual ground motion (cf. Fig. 2 of the main Letter), and its total thermal noise should be dominated by the pendulum mode from the TST stage. 

\begin{table}
 \caption{Summary of the LIGO-LF suspension parameters\label{tab:sus}}
 \begin{ruledtabular}
 \begin{tabular}{ccccc}
 Stage & mass [kg] & length [m] & Wire diameter [mm] & Material  \\
 \hline
TOP & 80 & 0.32 & 1.8 & C70 steel \\
UIM & 80 & 0.32 & 1.2 & C70 steel \\
PUM & 200 & 0.36 & 1.2 & Silica \\
TST & 200 & 1.2 & 0.6 (thin); 1.8 (thick) & Silica
 \end{tabular}
 \end{ruledtabular}
 \end{table}

To achieve the seismic isolation requirement, the mass ratio between the TOP and the TST stages should be similar to that of aLIGO. Decreasing the TOP mass shifts the highest suspension resonance to higher frequencies, making the pendulum filtering less efficient at 5\,Hz. Consequently we choose $m_{\rm TOP}=m_{\rm UIM} = 80\,{\rm kg}$, and the resultant seismic noise is shown in the dotted-brown curve of Fig. 1 of the main Letter. 

In addition to the direct length coupling, the longitudinal ground motion can also couple to the pitch motion of the test mass. The main pitch resonance frequency can be  controlled by tuning the distance between the fiber binding point and the mirror's center of mass. Similarly, the ground rotation can couple to the yaw motion of the test mass, and the resonance frequency can be controlled as well~\cite{Rakhmanov:00}. For LIGO-LF, the main pitch and yaw resonances are set to 0.42\,Hz and 0.35\,Hz, respectively, to balance the requirements for more filtering at high frequency ($>5\,{\rm Hz}$) and for less rms angular motions at low frequency ($<1\,{\rm Hz}$). 

\begin{figure}
 \includegraphics[width=0.48\textwidth]{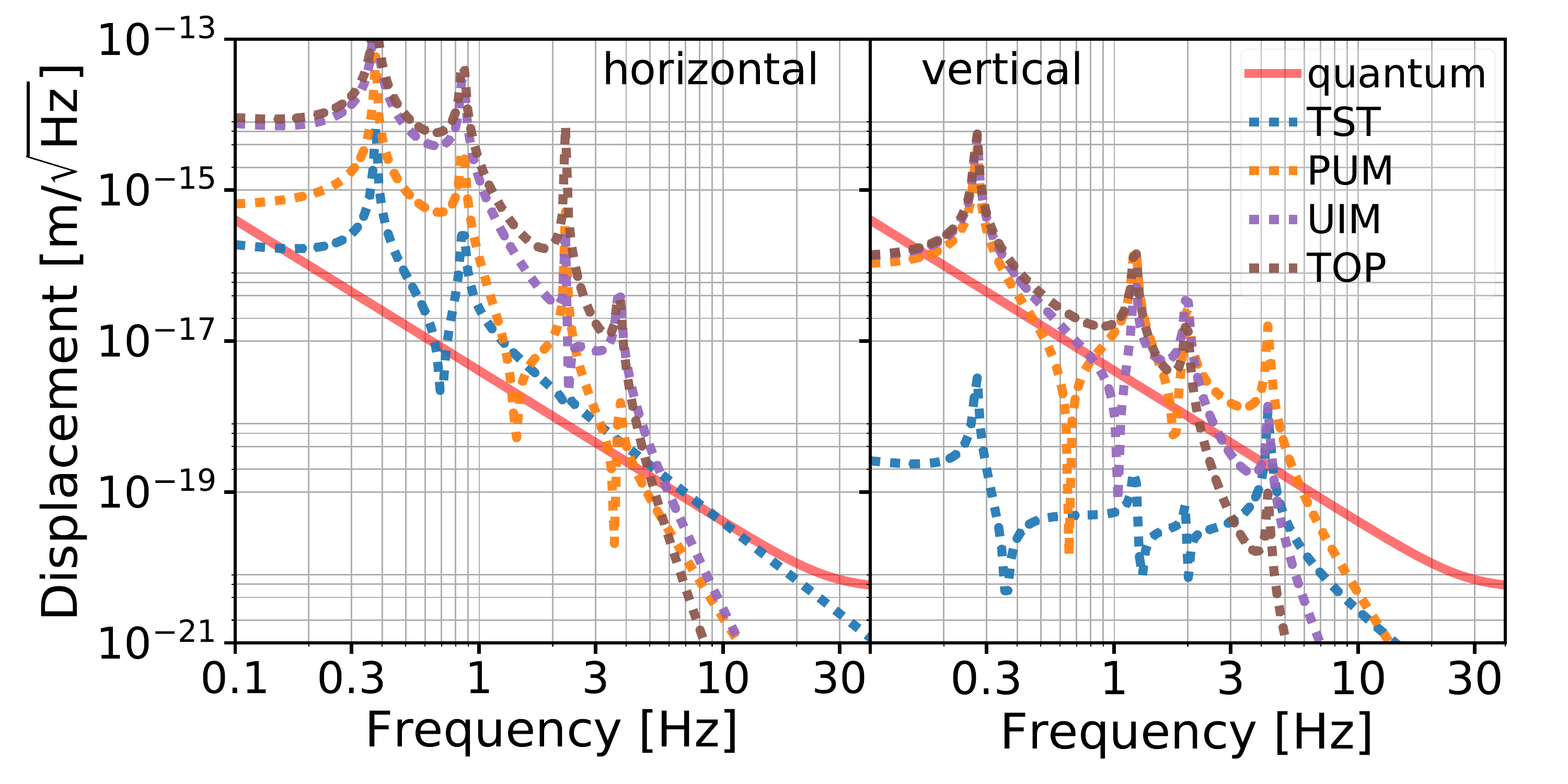}%
 \caption{The LIGO-LF suspension thermal noise from different stages (represented by dotted lines with different colors). The quantum noise is also plotted in the red-solid line as a reference. In the left we plot the direct horizontal (along the beam line) displacement noises. The dominant contribution above 5\,Hz is from the last stage  and it is similar to the quantum noise in the $5-20\,{\rm Hz}$ band.  In the right are the noises due to the vertical-to-horizontal coupling. The bounce mode is at $4.3\,{\rm Hz}$, making the vertical contributions subdominant above 5\,Hz.}
 \label{fig:sus_th}
\end{figure}

We present the suspension thermal noise for LIGO-LF in Fig. \ref{fig:sus_th}. In the sensitivity band above 5\,Hz, the dominant contribution comes from the pendulum mode of the test mass stage. In the calculation we have assumed an effective loss angle of $5\times10^{-10}$~\cite{Hammond:12} and the resultant suspension thermal noise is similar to the quantum noise from 5 to 20\,Hz. In order to reduce the contamination from other stages, we replace the suspension for the PUM stage from C70 steel wire to silica fiber. Meanwhile, the wire stress in the TOP and UIM stages is increased by $30\%$ relative to aLIGO for better dilution of the losses. 

Besides the thermal motion along the beam line, the vertical vibration of the test masses also couples to the GW channel due to the Earth's radius of curvature. 
The eigenfrequency $f_{\rm v}$ of the last stage's vertical mode (also known as the ``bounce mode'') scales as~\cite{Fritschel:02}
\begin{equation}
2\pi f_{\rm v} \approx \sqrt{\frac{gY}{l\sigma} \frac{m_{\rm TST} + m_{\rm PUM}}{m_{\rm PUM}}},
\end{equation}
where $g$, $Y$, $l$, $\sigma$, $m_{\rm TST}$, and $m_{\rm PUM}$ are the local gravitational acceleration, the Young's modulus of the material, the length of the suspension, the stress inside the fiber, the mass of the test mass, and the mass of the penultimate mass, respectively. 
To make $f_{\rm v}$ low, we maintain the mass ratio between the PUM and the test mass to 1 as aLIGO, and double $l$ to 1.2\,m. Meanwhile, the fibers suspending the test mass have a tapered geometry: for the thick part where most of the bending energy is stored, it has a diameter of $1.8\,{\rm \mu m}$ to cancel the thermal-elastic noise, while the thin part has a diameter of $0.6\,{\rm \mu m}$ to increase the stress $\sigma$ to 1.7\,GPa. Consequently, the bounce mode has an eigenfrequency of $f_{\rm v}=4.3\,{\rm Hz}$, which provides sufficient filtering of the vertical motion in the sensitivity band.

\section{Requirements of the angular control loops}\label{sec:asc}
The alignment loop of the arm cavity is designed to balance two requirements. On one hand, the loop needs to have large enough gain at low frequency to suppress the rms motion of the test mass to $\simeq 1\,{\rm nrad}$, and to overcome a radiation pressure induced angular instability. On the other hand, the loop gain needs to be as low as possible at high frequency to avoid perturbing the mirrors by feeding back the sensing noise. 

\begin{figure}
 \includegraphics[width=0.48\textwidth]{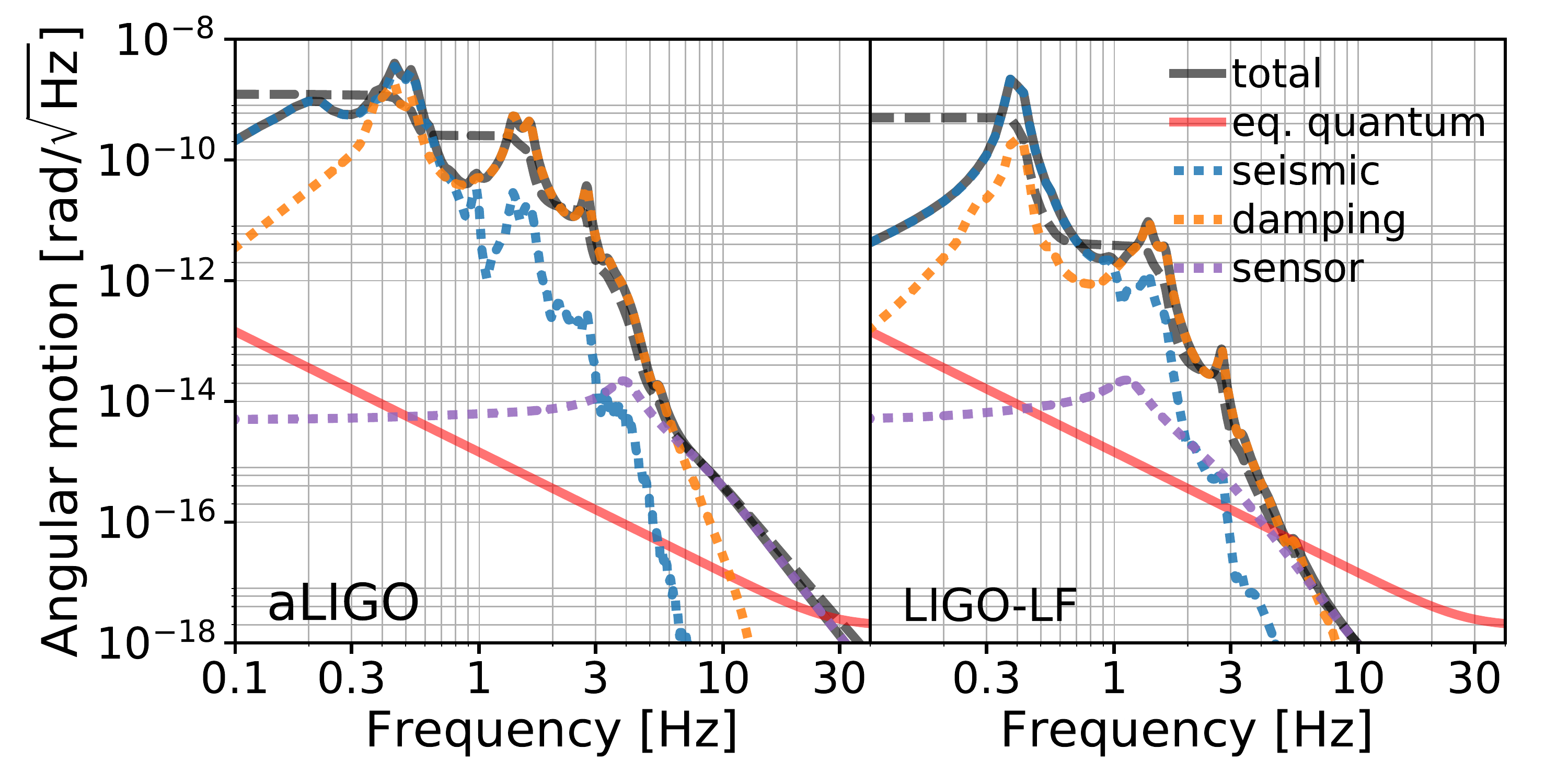}%
 \caption{The residual pitch motion of aLIGO (left panel) and LIGO-LF (right panel). The black-solid curves are the total angular motion and the black-dashed ones are the corresponding cumulative rms values. The dotted curves are the noise contributions due to seismic (blue), suspension damping (orange), and wave-front sensing (purple), respectively. The red-solid curve is shown for comparison: it corresponds to a noise level equivalent to the LIGO-LF's quantum noise if the spot miscentering is 1\,mm. }
 \label{fig:asc}
\end{figure}

In Fig. \ref{fig:asc} we plot the residual pitch motion for aLIGO and LIGO-LF after the alignment control is engaged with a detailed noise budget; the yaw motion is similar at high frequencies and is significantly less than pitch below 1\,Hz, so the low frequency rms requirement for yaw is less critical. In the calculation for aLIGO, we use the measured ground motion and shadow sensor noise to represent the contributions due to seismic and due to suspension damping, respectively. For LIGO-LF, we adopt the required sensor noise (the black trace in Fig. 2 of the main Letter) for the residual seismic motion, and scale the shadow sensor noise of aLIGO down by a factor of 100 for the damping noise. Therefore our results here should be interpreted as the requirement set for the future seismic  and damping sensors. The sensing noise from the wave-front sensors is assumed to be $5\times 10^{-15}\,{\rm rad/\sqrt{Hz}}$ for both aLIGO and LIGO-LF.  Also shown in the red curve as a comparison is the equivalent quantum noise: with 1\,mm of spot miscentering, an angular fluctuation per test mass given by the red curve will be converted to a length noise equal to the LIGO-LF's quantum limit. 

For aLIGO, a control bandwidth of 3\,Hz is necessary to reduce the rms pitch motion to $1.2\, {\rm nrad}$. Such a high bandwidth limits how fast the loop gain can roll-off at high frequency. Consequently, a considerable amount of control noise is injected to the 10-20\,Hz band, contaminating the GW sensitivity. For LIGO-LF, however, we can reduce the bandwidth to 0.8\,Hz, yielding a rms motion of $0.5\,{\rm nrad}$. We require the rms motion to be less than half of the aLIGO's value to open up the possibility of increasing the spot size by $50\%$. The sensing noise can now be decreased below the quantum limit at 4\,Hz. The damping noise, nonetheless, becomes significant for LIGO-LF, and a factor of 100 improvement is essential for reaching the instrument's fundamental limit at 5\,Hz. 

In addition to suppressing the test masses' rms motion, overcoming the angular instability is another critical requirement on the alignment bandwidth. With 0.8\,MW of power circulating in the arms, the radiation pressure force creates an optical torque comparable to the restoring torque from the suspension, and thus modifies the test masses' mechanical response. The input and end test masses are coupled by this effect to oscillate in a set of eigenmodes which are conventionally known as the ``hard'' and the ``soft'' modes~\cite{Sidles:06}.  Their eigenfrequencies are given by~\cite{Hirose:10}.
\begin{equation}
\omega_{\pm}^2=\omega_0^2 + \frac{PL}{Ic}\left[\frac{-(g_{\rm i} + g_{\rm e})\pm\sqrt{4+(g_{\rm i}-g_{\rm e})^2}}{1-g_{\rm i}g_{\rm e}}\right],
\end{equation}
where we have used $\omega_+$ ($\omega_-$) to represent the angular eigenfrequency of the hard (soft) mode, and $\omega_0$ the pendulum frequency. The $L$, $P$,  $I$, and $g_{\rm i\, (e)}$ are the arm length, power circulating in the arms, momentum of inertia, and the $g$ parameter of the input (end) test mass, respectively. The values for aLIGO are given in Ref.~\cite{Barsotti:10}, which leads to $\left(\omega_{-}/2\pi\right)^2=-(0.2\,{\rm Hz})^2$. The soft mode is thus unstable for aLIGO without control loop. Overcoming the instability will demand a bandwidth of $\gtrsim 10\,|\omega_-/2\pi|=2\,{\rm Hz}$. Nevertheless, as we increase the test masses by a factor of 5 to 200\,kg, the momentum of inertia will increase by a factor of $5^{5/3}\simeq15$ if we assume the mirror geometry stays the same as that of the aLIGO mirror. This increased momentum of inertia greatly suppresses the radiation pressure effect. Also taking into account the facts that we shift the pendulum frequencies for LIGO-LF (cf. Section \ref{sec:sus}) and modified the input test masses' radius of curvature to increase the spot size ($g_{\rm i}=-1.2$ for LIGO-LF; $g_{\rm e}$ is the same for LIGO-LF and aLIGO), the eigenfrequency of LIGO-LF's pitch (yaw) soft mode becomes $\left(\omega_{-}/2\pi\right) \simeq 0.4\,{\rm Hz}\ (0.3\,{\rm Hz})$. Consequently, the soft mode will be stable for LIGO-LF,  relaxing requirements set on the minimum control bandwidth.

\section{Calculation of the scattering noise}\label{sec:scat}
For the scattering noise calculation, we introduce the effective displacement $\bar{x}_{\rm scatter}$ defined as
\begin{equation}
\bar{x}_{\rm scatter}(t) = \frac{\lambda}{4\pi} \sin\left[ \frac{4\pi}{\lambda}x_{\rm scatter} (t) \right],
\end{equation}
where $x_{\rm scatter} (t)$ is the (physical) relative displacement between a mirror and a scattering surface at time $t$, and $\lambda = 1064\,{\rm nm}$ the laser wavelength. The corresponding frequency-domain displacement is thus given by
\begin{equation}
\hat{\bar{x}}_{\rm scatter} (f) = \frac{\lambda}{4\pi}\int \sin\left[ \frac{4\pi}{\lambda}x_{\rm scatter} (t) \right] \exp\left(-2\pi {\rm i} f t\right) {\rm d} t.
\end{equation}
Notice that when $x_{\rm scatter}\sim \lambda$, the effective displacement no more varies linearly with the physical displacement. Consequently, the large ground motion below 1\,Hz can be up-converted to the sensitivity band, making scattering a significant noise source when the ground motion is severe.  

\begin{figure}
 \includegraphics[width=0.48\textwidth]{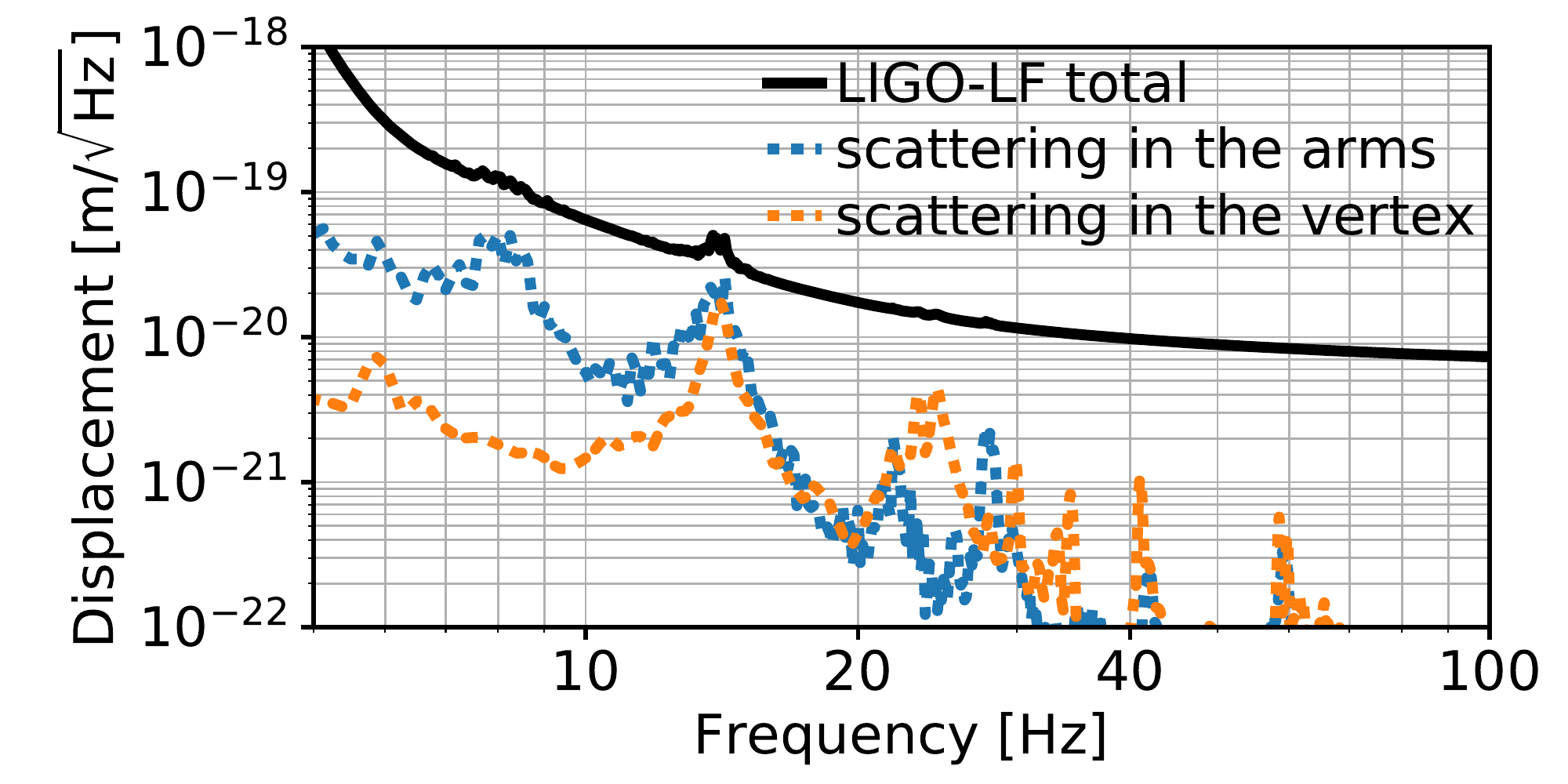}%
 \caption{The noises due to scattering in the arm tubes (dotted-blue) and in the vertex (dotted-orange). The total LIGO-LF noise is shown in the solid-black as a reference.}
 \label{fig:scatter}
\end{figure}

The olive trace in Fig. 1 of the main Letter is calculated including two effects: scattering in the arm tubes, and scattering in the vertex of the interferometer, with each one's contribution individually shown in Fig. \ref{fig:scatter}. 

For the former, the calculation follows from Ref.~\cite{Ottaway:12}. There are two coupling channels need to be considered. The phase quadrature of the scattered light directly enters the GW readout, with a flat transfer function from $\hat{\bar{x}}_{\rm scatter} (f)$ to the differential arm displacement given by
\begin{equation}
Z_{\rm tube}^{\rm (Im)}(f) \simeq 1.0\times10^{-12}\left(\frac{A_{\rm scat}}{2.0\times10^{-12}}\right)\,\frac{{\rm m}}{{\rm m}}.
\end{equation}
Here $A_{\rm scat}$ is the amplitude scattering coefficient and is proportional to the bi-directional reflectivity distribution function of the test masses~\cite{Flanagan:94}. At the same time, the amplitude quadrature of the scattered light can beat with the static optical field inside each arm to cause a differential power fluctuation, which further induces differential arm motions due to the radiation pressure force. The transfer function for this mechanism is given by
\begin{align}
Z_{\rm tube}^{\rm (Re)}(f) &\simeq 3.3\times 10^{-12} \left(\frac{10\,{\rm Hz}}{f}\right)^2\left(\frac{A_{\rm scat}}{2.0\times10^{-12}}\right)\nonumber \\
&\times\left(\frac{200\,{\rm kg}}{m_{\rm TST}}\right)\left(\frac{P_{\rm arm}}{0.8\,{\rm MW}}\right)\left(\frac{T_{\rm srm}}{0.25}\right)\, \frac{{\rm m}}{{\rm m}},
\end{align}
where $P_{\rm arm}$ is the power circulating in each arm and $T_{\rm srm}$ is the power transmissivity of the signal recycling mirror. Below 20\,Hz, the amplitude quadrature dominates the coupling. 

The scattering in the vertex is caused by the anti-reflecting (AR) surfaces along the optical path. If not properly baffled, the stray light may hit the chamber wall and  be reflected back to the optical path. The coupling coefficient per stray beam is~\cite{llo:29665}
\begin{align}
Z_{\rm vertex}(f) &\simeq 1.0 \times 10^{-12} \left(\frac{T_{\rm baffle}}{0.001}\right)^{1/2}\nonumber\\
&\times \left(\frac{R_{\rm AR}}{250\,{\rm ppm}}\right)\left(\frac{2\,{\rm mm}}{w_{\rm wall}}\right)\, \frac{{\rm m}}{{\rm m}},
\end{align}
where $R_{\rm AR}$ is the power reflectivity of the AR surface creating the beam, $w_{\rm wall}$ is the stray light's spot size on the chamber wall, and $T_{\rm baffle}$ is the fraction (in power) of the stray light that leaks through the baffle. There are 10 AR surfaces that can contribute to this noise, 2 from the input test masses ($R_{\rm AR}\simeq 250\,{\rm ppm}$), 4 from the beam splitter ($R_{\rm AR}\simeq 50\,{\rm ppm}$), and 4 from the compensation plates ($R_{\rm AR}\simeq20\,{\rm ppm}$).  To achieve the proposed LIGO-LF sensitivity, the baffles need to reduce the power of the stray light by
\begin{equation}
T_{\rm baffle}^{\rm (LIGO-LF\ req.)}<0.1 \%.
\end{equation}

\section{Estimation of detection rate}\label{sec:det_rate}
We present the our calculation of the expected detection rate of merging binary BHs in this section. 

We adopt the standard power-law mass distribution used in LIGO event rate estimation~\cite{O1BBH:16}. The probability densities of the primary mass $M_1$ and the secondary mass $M_2$ are respectively given by
\begin{align}
&\mathcal{P}_1(M_1) = \mathcal{A}_{M_1} M_1^{-\alpha}\Theta\left(M_1-M_{\rm gap}\right)\exp\left(-\frac{M_1}{M_{\rm cap}}\right), \\
&\mathcal{P}_2(M_2) = \mathcal{A}_{M_2} \Theta\left(M_2-M_{\rm gap}\right)\Theta\left(M_1-M_2\right),
\end{align}
where $A_{M_1}$ and $A_{M_2}$ are overall normalizations, and $\Theta$ denotes the Heaviside step function. Following the convention we use a slope of $\alpha=2.35$ and a lower limit of the mass distribution $M_{\rm gap}=5\,M_\odot$. As in Ref.~\cite{Ely:17}, we have an exponential cutoff on $M_1$ which is set to $M_{\rm cap}=60\,M_\odot$. Additionally we require $M_1+M_2\le100\,M_\odot$. We do not consider the IMBHs in our rate calculation because of the large uncertainty in their formation; they are sufficiently rare and are thus unlikely to affect the total number of detections. For the merging rate, we adopt a simple, mass-independent approximation~\cite{Ely:17} 
\begin{equation}
\mathcal{R}(z)=97(1+z)^2\,{\rm Gpc^{-3}\,yr^{-1}}. 
\end{equation}

The expected number of detection per unit time $\diff T$ in the mass interval $\left[M_{\rm tot}, M_{\rm tot} + \diff M_{\rm tot}\right]$ is given by
\begin{align}
\frac{\diff N\left(M_{\rm tot}\right)}{\diff T\diff M_{\rm tot}}=&4\pi\int_{M_{\rm tot}/2}^{M_{\rm tot}}\diff M_1 \mathcal{P}_1\left(M_1\right)\mathcal{P}_2\left(M_{\rm tot}-M_1\right) \nonumber \\
&\times\int_{0}^{z_{\rm ran}(M_1, M_{\rm tot}-M_1)}\frac{c\chi(z)^2\mathcal{R}(z)}{(1+z)H(z)} \diff z,
\end{align}
where $z_{\rm ran}(M_1, M_2)$ is the detection range for a binary system with $M_1$ and $M_2$, $\chi(z)$ the radial comoving distance, and $H(z)$ the Hubble parameter. In the calculation we have assumed a flat universe with Hubble constant $H_0=67\,{\rm km\,s^{-1}\,Mpc^{-1}}$ and matter (dark energy) fraction $\Omega_{m}=0.32$ ($\Omega_\Lambda = 0.68$), consistent with the Planck result~\cite{Planck:16}. The $z_{\rm ran}(M_1, M_2)$ is calculated with a single detector when we do the rate estimation. This is because we would like to focus on the improvements due to better sensitivity, instead of due to more detectors or more optimized network configuration. 

We divide the total mass $M_{\rm tot}$ into 8 logarithmically spaced bins from $10\,M_\odot$ to $100\,M_\odot$. For the $i$th bin, we compute the quantity $\Delta N^i=\left\{\left[\diff N (M_{\rm tot}^i)/\diff T \diff M_{\rm tot}\right]T_{\rm 1yr}\,\Delta M_{\rm tot}^i\right\}$, where $T_{1yr}$ is the time for a year and $\Delta M_{\rm tot}^i$ is the width of the $i$th bin. 

The result is shown in the left panel of Fig. 3 in the main Letter. Note that $\Delta N^i$ yields a Poisson distribution, with a statistical uncertainty of $\sqrt{\Delta N^i}$. Therefore the statistical SNR grows as $\sqrt{\Delta N^i}$, greatly enhancing LIGO-LF's ability to constrain the population properties of binary BHs relative to aLIGO. Consider a simple case where the merging rate $\mathcal{R}$ is the only unknown, then with LIGO-LF we would be able to constrain it to within $\pm 0.5 \,{\rm Gpc^{-3}\,yr^{-1}}$ by summing all the mass bins together for a total observation period of 10-year, 4 times better than what aLIGO can achieve. As $\mathcal{R}$ is sensitive to, e.g., the metallicity at the time of binary formation~\cite{2016ApJ...818L..22A},  an accurate measurement of  $\mathcal{R}$ thus constrains the metal enrichment history of the Universe. Furthermore, the event rate per mass interval can also be used to place constraints on the fraction of dark matter in the Universe that is in the form of primordial BHs~\cite{Bird:16, Ely:17}.

\section{Inferred masses in the source frame}\label{sec:pe}

In the main Letter, the parameter estimation section focused on the results in the detector-frame. Here we provide the source-frame results for completeness. The conversion is~\cite{Cutler:94}
\begin{equation}
\mathcal{M}_c = \frac{\mathcal{M}_c^{\rm (d)}}{1+z}
\end{equation}
for the chirp mass $\mathcal{M}_c$, and similarly for the total mass $M_{\rm tot}$. Here $z$ is the cosmological redshift and we have denoted the detector-frame with a superscript (d). 

\begin{figure}
 \includegraphics[width=0.48\textwidth]{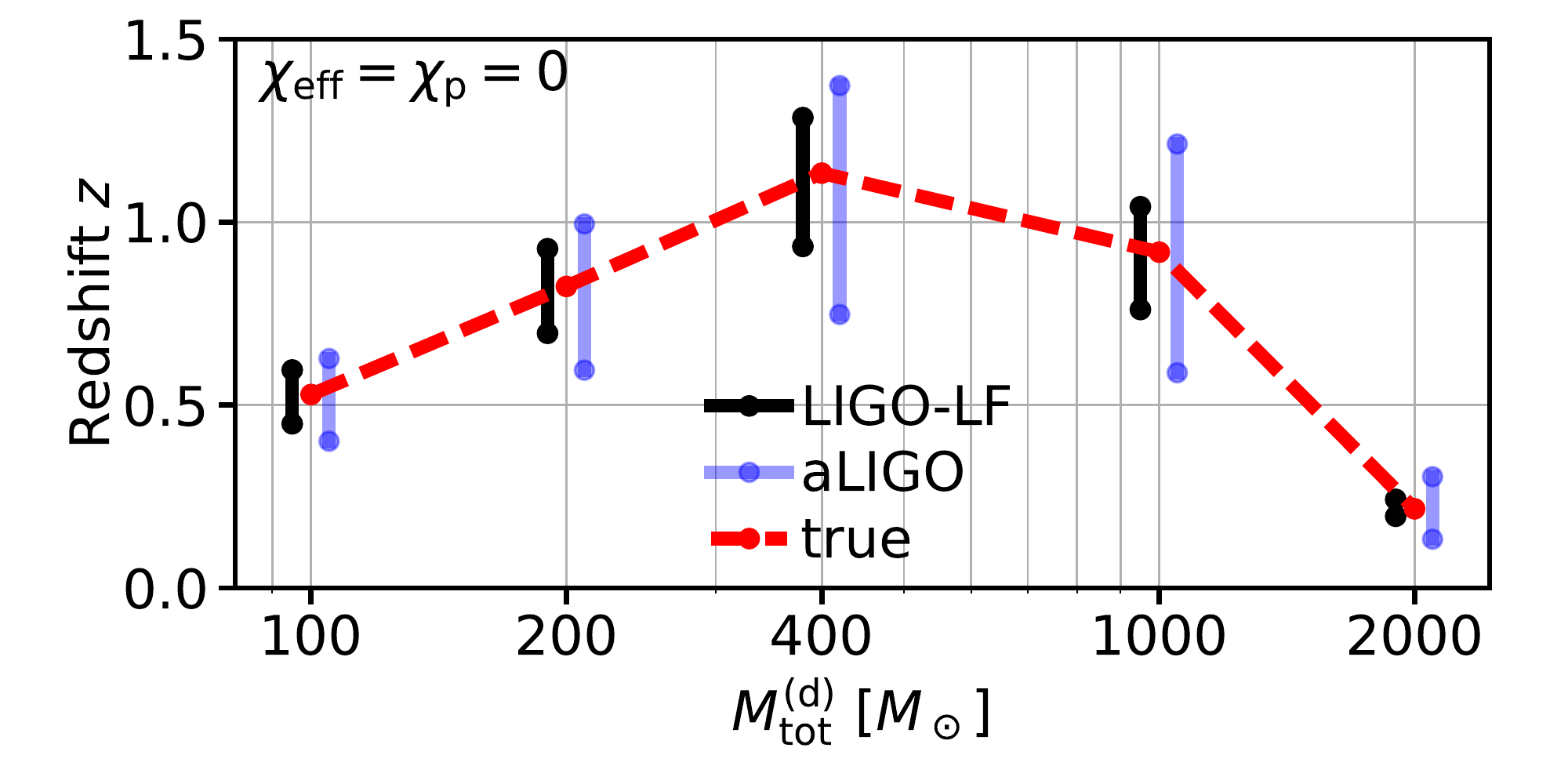}%
 \caption{Mock sources for each total mass were placed at the redshifts indicated by the red-dashed line. The redshifts were chosen to give a network SNR of 16 in aLIGO. The black (blue) bars indicate the $90\%$ credible interval for the inferred redshift with LIGO-LF (aLIGO) sensitivity.  LIGO-LF typically improves the constrain in $z$ by a factor of 2. 
\label{fig:redshift}}
\end{figure}

In Fig. \ref{fig:redshift} we present the $90\%$ credible intervals of the redshift $z$. To yield a network SNR of 16 with aLIGO design sensitivity, the redshifts are $z=\left(0.53,\ 0.82,\ 1.1,\ 0.92,\ 0.22\right)$ for the 5 injections we have with $M_{\rm tot}^{\rm (d)} = \left(100,\ 200,\ 400,\ 1000,\ 2000\right)\,M_\odot$, respectively.     LIGO-LF typically improves the accuracy in the redshift inference by a factor of 2 relative to aLIGO. 

\begin{figure}
 \includegraphics[width=0.48\textwidth]{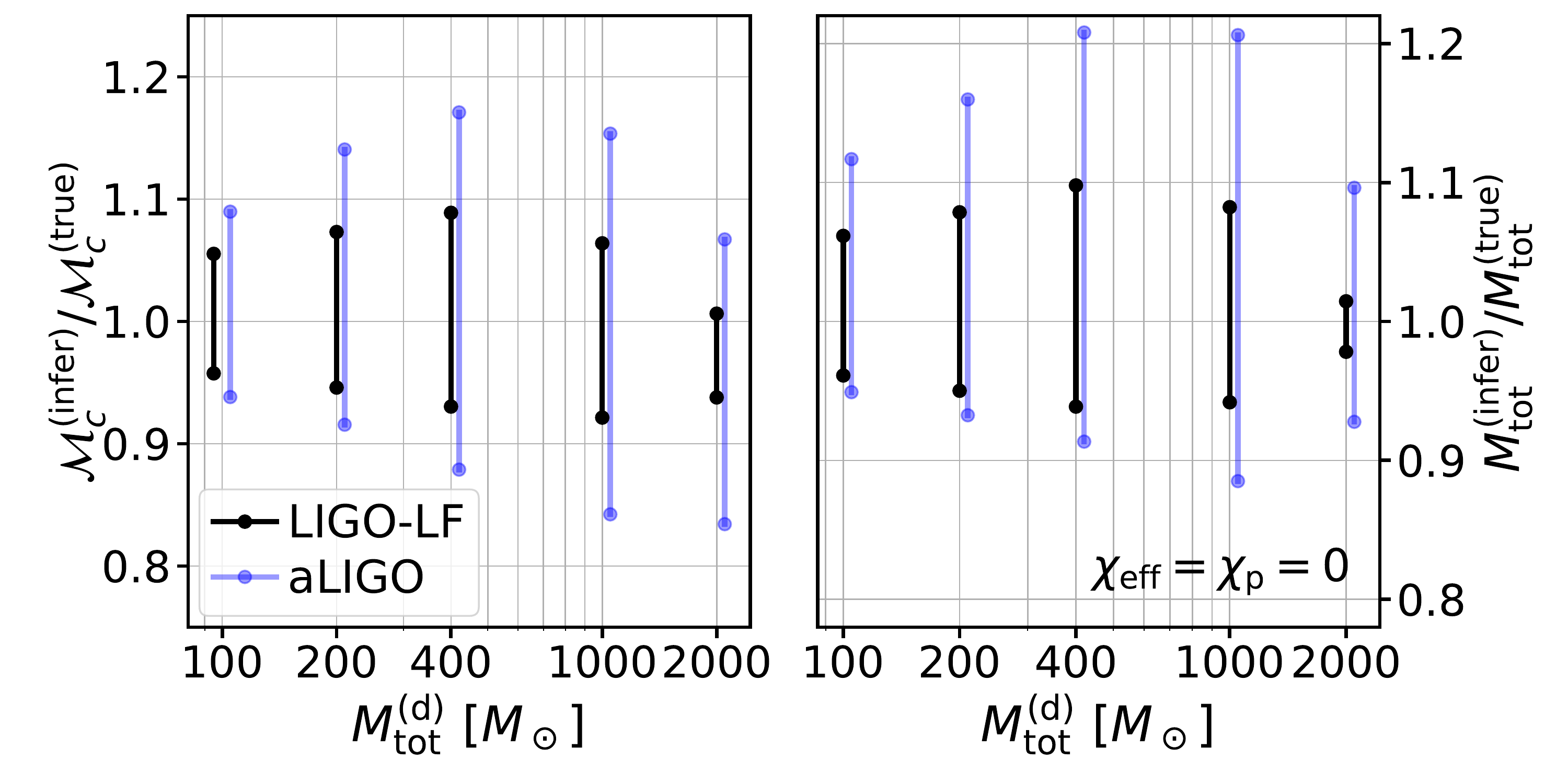}%
 \caption{The $90\%$ credible interval of the source-frame chirp mass $\mathcal{M}_c$ (left panel) and total mass $M_{\rm tot}$ (right panel). The uncertainty is about a factor of 2 smaller for LIGO-LF compared to aLIGO, and is dominated by the uncertainty in inferring the redshift. Thus LIGO-LF enables less improvement in constraining the source-frame masses than in the detector-frame ones.  \label{fig:PE_mass_source_q1}}
\end{figure}

We show the $90\%$ credible intervals of the source-frame masses in Fig. \ref{fig:PE_mass_source_q1}. The injected source-frame total masses are $M_{\rm tot}=\left(65, 109, 187, 521, 1644\right)\,M_\odot$, and chirp masses $\mathcal{M}_c = \left(28,\ 48,\ 82,\ 227,\ 716\right)\,M_\odot$. Due to the statistical error in measuring the redshift, LIGO-LF only constrains the source-frame values 2 times better than aLIGO, despite that it can constrain the detector-frame ones 3-5 times better. 

\section{Localization of merging binary NSs}\label{sec:bns_loc}
We consider localizing a merging binary NS with a network of 4 detectors consisting of the Hanford (H) and the Livingston (L) sites, LIGO-India (I), and Virgo (V). The coordinates of HLV can be found in Ref.~\cite{Jaranowski:98}, and we use the same location for I as in Ref.~\cite{Vitale:17}. The data seen in the $i$th detector in the network can be written as~\cite{Zhao:17, Wen:10}
\begin{equation}
d_i(f)=\left[F_i^+(f)h_+(f) + F_i^\times(f) h_\times (f)\right]\exp\left[-\imag 2\pi f \tau_i(f)\right],
\end{equation}
where $h_{+(\times)}$ is the GW signal in the plus (cross) polarization, $F_{+(\times)}$ the antenna response whose functional form is provided in~\cite{Jaranowski:98}, and $\tau_i$ the traveling time from the coordinate origin to the $i$th detector. The frequency dependences of $F_{+(\times)}$ and $\tau_i$ originate from the rotation of the Earth, and with the stationary-phase approximation they can be written as,
\begin{align}
&F_{+(\times)}(f)=F_{+(\times)}\left[t_f(f)\right],\hspace{0.5cm}\tau_i(f)=\tau_i\left[t_f(f)\right],\\
&t_f=t_c-\frac{5}{256}\left(\frac{G\mathcal{M}_c}{c^3}\right)^{-5/3}\left(\pi f\right)^{-8/3}.
\end{align}
Here $t_c$ is the time of the coalescence. We compute the signal $h(f)$ using the post-Newtonian (PN) expansion, including  phase corrections to the 1.5 PN order~\cite{Cutler:94}. 

We parameterize the signal in terms of 9 parameters: $\vect{p}=\left(\mathcal{M}_c,\ q,\ t_c,\ \phi_c,\ \iota,\ \theta_s,\ \phi_s,\ \psi_s,\ d_{\rm L}\right)$, corresponding to the chirp mass, mass ratio ($\le 1$), time and phase at the merger, the source's inclination, declination, right ascension, polarization, and luminosity distance. The spin is not included since the NS is expected to be slow spinning~\cite{GW170817, Burgay:03}. The statistical error of each parameter can be estimated using the Fisher matrix with element
\begin{equation}
\Gamma_{jk}=\left\langle\frac{\partial \vect{d}}{\partial p_j},\,\frac{\partial{\vect{d}}}{\partial p_k}\right\rangle,
\end{equation}
where the inner product for data from the network, $\vect{a}$ and $\vect{b}$, is defined as 
\begin{equation}
\left\langle\vect{a}, \vect{b}\right\rangle(f_{\rm up})=2\sum_i^{\rm HLVI}\int_0^{f_{\rm up}}\diff f\left[\frac{a_i^\ast(f) b_i(f)+a_i(f) b_i^\ast(f)}{S_i(f)}\right]. \label{eq:inner_prod}
\end{equation}
Here $S_i(f)$ is the noise power spectra density of the $i$th detector, $f_{\rm up}\le 2f_{\rm ISCO}$ is the upper limit of the integration, and $f_{\rm ISCO}$ is the orbital frequency at the system's inner-most stable circular orbit (ISCO). We have treated $f_{\rm up}$ as a free parameter so that we can consider the cumulative accuracy using only data with $f<f_{\rm up}$, 

The full covariance matrix $\vect{\Sigma}$ can be obtained by inverting $\vect{\Gamma}$,
\begin{equation}
\vect{\Sigma}=\vect{\Gamma}^{-1},
\end{equation}
and the statistical error for the $j$th parameter $p_j$ is given by 
\begin{equation}
\Delta p_j=\sqrt{\Sigma_{jj}}.
\end{equation}
We are especially interested in the uncertainty solid angle covering the source's location, which is given by
\begin{equation}
\Delta \Omega_s = 2\pi |\sin \theta_s| \sqrt{\left\langle \Delta \theta_s^2\right\rangle\left\langle \Delta \phi_s^2\right\rangle-\left\langle \Delta \theta_s\Delta \phi_s\right\rangle^2}.
\end{equation}

In order to demonstrate the improvement made by LIGO-LF over aLIGO and A+, we focus on $1.4\,M_\odot$-$1.4\,M_\odot$ NS binaries, and fix the source location to the Coma cluster. We consider two inclination angles, a face-on one with $\iota = 30^{\circ}$, and a more edge-on one with $\iota = 75^{\circ}$. The arriving time and polarization of the sources are marginalized over. 

In the left panel of Fig. 5 in the main Letter, we plot the cumulative localization error, $\Delta \Omega_s(f_{\rm up})$. Here instead of integrating eq.~(\ref{eq:inner_prod}) over the entire band, we integrate it only up to $f_{\rm up}$. We can thus know the localization accuracy at each instant of the inspiral. As shown in the figure, LIFO-LF localize the source 5 (10) times better than A+ (aLIGO) at 30\,Hz, and 10 (15) times better at 20\,Hz. Note that the time prior the final merger increases sharply as the frequency decreases, as
\begin{equation}
t_{c}-t_f=54\left(\frac{\mathcal{M}_c}{1.2\,M_\odot}\right)^{-5/3}\left(\frac{f}{30\,{\rm Hz}}\right)^{-8/3}\,{\rm s}.
\end{equation}
Despite that the final uncertainties are similar for A+ and for LIGO-LF, LIGO-LF would be able to send out the source location minutes before the final merger, while for A+ similar accuracy cannot be achieved until seconds before the event. 

\section{The detectability of the NS $r$-mode resonances}

The (linear) tidal response of the NS can be decomposed into an equilibrium tide and a dynamical tide. The equilibrium tide accounts for the quasi-static, large-scale distortion of the star and the dynamical tide~\cite{Lai:94, Reisenegger:94} accounts for the internal modes of oscillation that are resonantly excited as the orbit decays and sweeps up in frequency. Here we consider the excitation of the NS's rotational modes (i.e. the $r$-modes) due to its companion's gravitomagnetic tidal field~\cite{Flanagan:07}.  For the $l=2,\,m=1$ $r$-mode we study here, the GW frequency of mode resonance $f_r$ is related to the NS's spin frequency $f_{\rm spin}$ as
\begin{equation}
f_{r}=\frac{4}{3}f_{\rm spin}. \label{eq:rmode_freq}
\end{equation}
Since the NSs in binary NS systems are expected to be slow-spinning (with a rate less than a few$\times10\,{\rm Hz}$), the $r$-mode is naturally an interesting science case for the LIGO-LF upgrade. 

The tidal interaction induces a phase shift in the GW, $\delta \Phi_r$, relative to the point-particle (pp) waveform. As the duration of the mode resonance is typically $\sim 1\%$ of the total GW decay timescale~\cite{Flanagan:07}, the resonance can thus be treated as an instantaneous process. In this limit, the phase of the frequency-domain waveform $\Psi(f)$ can be written as~\cite{Yu:17b} 
\begin{equation}
\Psi(f)=\Psi_{\rm pp}(f) - \left(1-\frac{f}{f_r}\right)\delta \Phi_r \Theta\left(f_r-f\right), \label{eq:tidal_shift}
\end{equation}
where $\Psi_{\rm pp}$ is the phase of the point particle waveform (calculated to the 1.5 PN order in our study). In the expression above we have aligned the tidal waveform to the pp one at the merger. 

While in the case of equilibrium tides the orbital energy is absorbed by the star and thus the inspiral is accelerated, in the case of the $r$-mode interaction the direction of energy transfer is reversed. The orbit extracts the NS spin energy which decelerates the inspiral. This unique feature of the $r$-mode corresponds to a \emph{negative} $\delta \Phi_r$ in the expression above. While large theoretical uncertainties exist, previous work suggests that~\cite{Flanagan:07}
\begin{equation}
\delta \Phi_r \sim -0.1 \left(\frac{f_{\rm spin}}{100\,{\rm Hz}}\right)^{2/3}. \label{eq:rmode_amp}
\end{equation}

To estimate the detectability of this effect, we once again use the Fisher matrix method. A fully Bayesian analysis is deferred to future studies. For simplicity, we focus on the single-detector case and fix the sources at $50\,{\rm Mpc}$ with optimal orientation. This allows us to write the waveform in terms of 9 parameters, $\left(\mathcal{M}_c,\ q,\ \chi_1,\ \chi_2,\ t_c,\ \phi_c,\ d_{\rm L},\  f_r,\ \delta\Phi_r\right)$, corresponding to the chirp mass, mass ratio, dimensionless spin of mass 1 and 2, time and phase at the coalescence, luminosity distance, and the resonant frequency and the phase shift of the $r$-mode. The equilibrium tidal deformation is not included here because it is only relevant at $f\gtrsim600\,{\rm Hz}$~\cite{Hinderer:10}; it is unlikely for a NS to spin this fast. We consider here binaries with $M_1=1.4\,M_\odot$ and $M_2=1.35\,M_\odot$. The mass ratio is slightly off 1 because otherwise $\chi_1$ and $\chi_2$ will be completely degenerate. The relation between the spin frequency $f_{\rm spin}$ and the dimensionless spin parameter $\chi$ depends on the NS EOS. Here we pick the SLy EOS~\cite{Douchin:01} as a typical representation; this EOS is consistent with the GW170817 event~\cite{GW170817}. It leads to
\begin{equation}
\chi\simeq0.06\left(\frac{f_{\rm spin}}{100\,{\rm Hz}}\right),\label{eq:fs2chi}
\end{equation}
for a typical $1.4\,M_{\odot}$ NS. A softer EOS yields a larger $\chi$ for a NS with fixed mass. 

We then vary the spins of the two masses (while keeping them rotating at the same rate as $f_{\rm spin 1}=f_{\rm spin 2}=f_{\rm spin}$), and evaluate the Fisher matrix at different values of $f_{\rm spin}$ to obtain the uncertainty on the phase shift, $\Delta \left(\delta \Phi_r\right)$. 

The results plotted in the right panel of Fig. 5 in the main Letter. The $r$-mode is detectable if the statistical error is smaller than the real phase shift calculated from eq. (\ref{eq:rmode_amp}), i.e., we set $\Delta \left(\delta \Phi_r\right)\le|\delta \Phi_r|$ as the detectability threshold. 

Note that we included only a single set of $\left(f_r,\ \delta \Phi_r\right)$ in the calculation above, whereas in a merging binary NS system each NS should contribute individually a $r$-mode phase shift. Nevertheless, the typical resolution of $f_r$ is $\simeq 50\,{\rm Hz}$, corresponding to a spin frequency of $\Delta f_{\rm spin}\simeq 40\,{\rm Hz}$, whereas even the fastest spinning pulsar in a binary NS system known today~\cite{Burgay:03} will have a rotation rate less than that when it enters the sensitivity band of a ground-based GW detector. We are thus unlikely to resolve the individual resonance but only the combined effect of the two NSs. Therefore we included an extra factor of 2 when computing the theoretical prediction (i.e. the red-dashed curve) according to eq. (\ref{eq:rmode_amp}). 

We also point out that this is not the only way to parameterize the $r$-mode resonance. For completeness we introduce an alternative parameterization which involves a single parameter $A_r$ controlling the overall coupling strength of the $r$-mode. Instead of treating the resonant frequency as a free parameter, we utilize eq. (\ref{eq:rmode_freq}) to relate $f_{r1}$ to $f_{\rm spin1}$, which is further related to $\chi_1$ by eq. (\ref{eq:fs2chi}) (by doing so we implicitly restrict ourselves to a fixed EOS). The phase shift can then be calculated as~\cite{Flanagan:07}
\begin{align}
\delta \Phi_{r1}&=-A_r\left(\frac{f_{\rm spin}}{100\,{\rm Hz}}\right)^{2/3}\nonumber \\
	&\times\left(\frac{1.4\,M_\odot}{M_1}\right)\left(\frac{1.4\,M_\odot}{M_2}\right)^2\left(\frac{2.8\,M_\odot}{M_{\rm tot}}\right)^{1/3}\,{\rm rad},\label{eq:rmode_alt}
\end{align}
for the primary mass, and switching the index $1\leftrightarrow2$ for the secondary mass. Both $\delta \Phi_{r1}$ and $\delta \Phi_{r2}$ are then included to the waveform according to eq. (\ref{eq:tidal_shift}). 

\begin{figure}
 \includegraphics[width=0.48\textwidth]{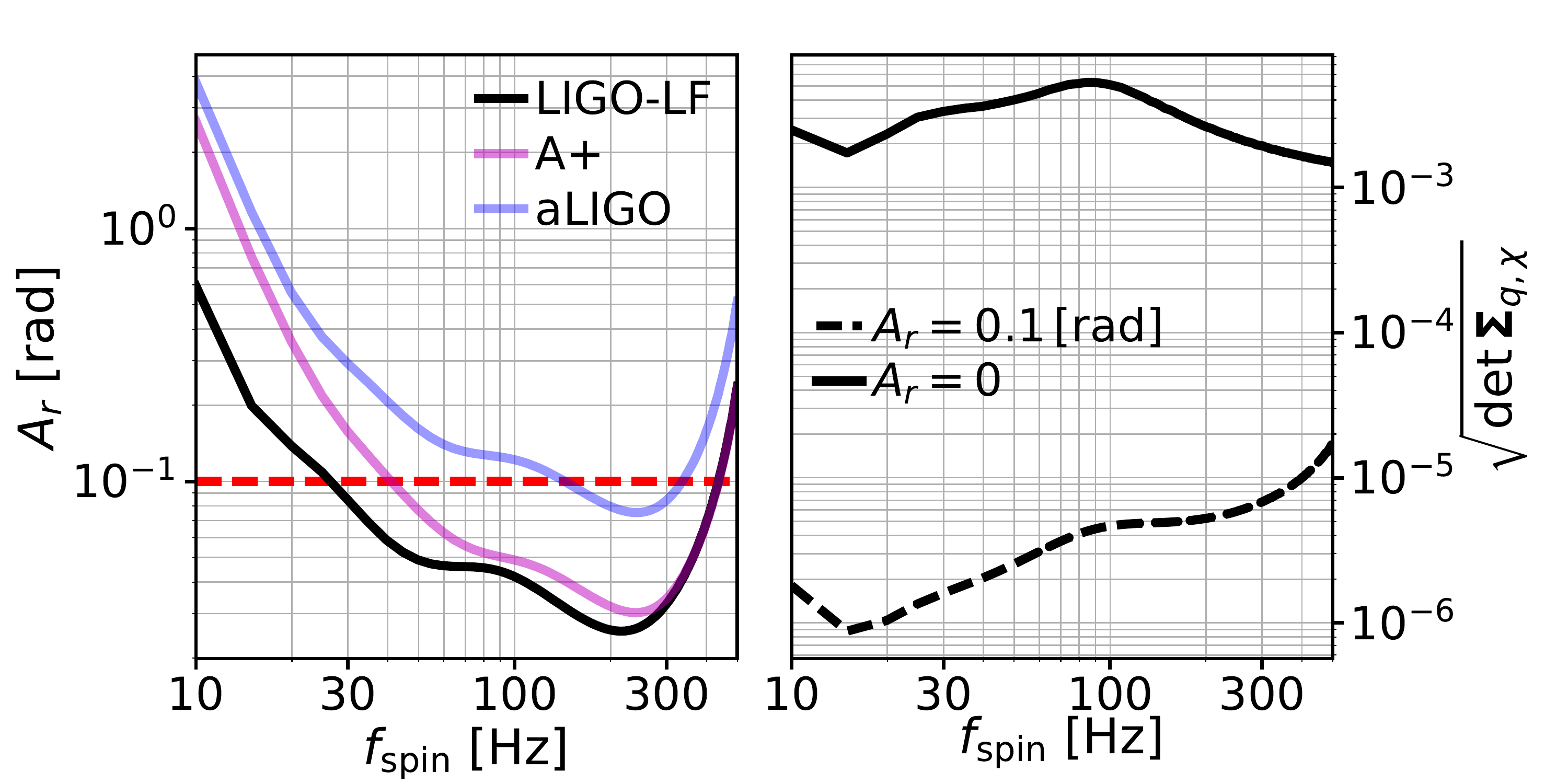}%
 \caption{Left: the parameter estimation uncertainty (solid lines) of the $r$-mode phase shift with an alternative parameterization (cf. eq.~\ref{eq:rmode_alt}). The expected value of $A_r$ is also shown in the red-dashed line. The $r$-mode is detectable when the dashed line is above the solid line. The detectable region is consistent with the one obtained with the parameterization we used in the main Letter. 
 Right: the volume of the error ellipsoid pertaining $\left(q,\ \chi_1,\ \chi_2\right)$ for the same sources and with the LIGO-LF sensitivity. When the $r$-mode effect is included (dashed), the error ellipsoid has a significantly smaller volume than the one without considering the $r$-mode effect (solid). 
 \label{fig:rmode_supp}}
\end{figure}

This allows us to write a Fisher matrix with 8 parameters, $\left(\mathcal{M}_c,\ q,\ \chi_1,\ \chi_2,\ t_c,\ \phi_c,\ d_{\rm L},\  A_r\right)$.  We consider the same sources as above and evaluate the matrix at different spin frequencies but still with $\chi_1=\chi_2$. We once again set the detectability threshold to $\Delta A_r\le A_r\sim0.1\,{\rm rad}$. The range of NS spin frequency in which the $r$-mode is detectable is comparable to the parameterization we adopt in the main Letter, as shown in Fig.~\ref{fig:rmode_supp}.

As pointed out in the main Letter and also in Ref.~\cite{Andersson:18}, though the equilibrium tides already place constraints on the NS EOS, the constraint relies only on the bulk property of the star. The dynamical tides such as the $r$-mode resonances, however, probe quantities like the internal composition of the NS, which may provide much detailed information about the EOS. 

Additionally, the confirmation of the $r$-mode may potentially improve the PE accuracy of other parameters. This effect is illustrated in the right panel of Fig. \ref{fig:rmode_supp}. Here we consider the same source as above and focus on cases with the LIGO-LF sensitivity. Specifically, we consider the error ellipsoid corresponding to $\left(q,\ \chi_1,\ \chi_2\right)$. The volume of this ellipsoid $\Delta V_{q,\chi}$ is given by
\begin{equation}
\Delta V_{q,\chi}=v_0\sqrt{\det\,\vect{\Sigma}_{q,\chi}},
\end{equation}
where $v_0$ is a geometrical constant, $\vect{\Sigma}$ is the covariance matrix obtained by inverting the Fisher matrix (using the alternative parameterization). Here we focus on the submatirx pertaining $\left(q,\ \chi_1,\ \chi_2\right)$, as denoted by the subscript $(q,\chi)$. When the $r$-mode is included, the volume of uncertainty decreases dramatically. This is due to the fact that in the pp waveform we can only measure $\chi_{\rm eff}$, which is approximately the average of $\chi_1$ and $\chi_2$ for binaries with nearly equal masses, and even $\chi_{\rm eff}$ is partially degenerate with the mass ratio $q$. When the $r$-mode is included, the frequencies of resonance directly measure the spin of each NS, which breaks the degeneracy between parameters. Once $\chi_1$, $\chi_2$ and $q$ are better constrained, the accuracy in measuring the tidal deformability due to the equilibrium tide can also be improved~\cite{GW170817}. 

\section{Searching for the GW memory effect}
The GW memory causes a DC displacement of the test masses that persists after the GW has passed~\cite{Christodoulou:91}. As the effect builds up over a finite amount of time, it can thus be detected by a LIGO-like GW detector which effectively high-passes the signal. The detection of this effect may provide a strong-field test of the general theory of relativity. Therefore we consider it as one of the science cases for LIGO-LF. 

Here we adopt the minimal-waveform model proposed in Ref.~\cite{Favata:09}, 
\begin{equation}
h_+^{\rm (mem)}(f)=\frac{G \eta M_{\rm tot}/c^2}{384\pi D_{\rm L}}\sin^2\iota\left(17+\cos^2\iota\right)h^{\rm (mem)}(f),
\end{equation}
where $\eta=M_1M_2/M_{\rm tot}^2$ is the symmetric mass ratio, and $h_\times^{\rm (mem)}(f)=0$. The term $h^{\rm (mem)}(f)$ is further given by
\begin{align}
h^{\rm (mem)}(f)=&\frac{\imag}{2\pi f}
\Big{\{}
\frac{8\pi G M_{\rm tot}}{c^2r_{\rm m}}\left[1-2\pi\imag f \tau_{\rm rr} U(1,7/4,2\pi\imag f \tau_{\rm rr})\right] \nonumber \\
&-\frac{c^3}{G\eta M_{\rm tot}}\sum_{n,n'}^{n_{\rm max}}\frac{\sigma_{22n}\sigma^\ast_{22n'}A_{22n}A^\ast_{22n'}}{2\pi\imag f - (\sigma_{22n}+\sigma_{22n'}^\ast)}
\Big{\}}.
\end{align}
The value of $\tau_{\rm rr}=\left(5/256\right)\left(GM_{\rm tot}/c^3\eta\right)\left(c^2r_{m}/GM_{\rm tot}\right)^4$ is the characteristic orbital decay time scale at $r_{\rm}$, and $U$ is Kummer's confluent hypergeometric function of the second kind. The $\sigma_{lmn}$ are angular frequencies of the final BH's quasi-normal modes, whose value are given in Ref.~\cite{Berti:06}. The coefficients of $A_{lmn}$ can be solved by matching the leading order quadrupole moments in the inspiral phase to the sum of the ringdown normal modes at $r_{\rm m}.$ Here we choose $r_{\rm m}$ to correspond to the orbital separation at the ISCO. 

In our calculation we consider a simple case where we fix the source distance to $z=0.1$ ($D_{\rm L}=0.48\,{\rm Gpc}$) and inclination to $\iota=30^{\circ}$. We further assume that the signal is purely in the ``$+$'' polarization. We then vary the source-frame total mass (while keeping the mass ratio to 1) and compute the single-detector matched-filter SNR for each source with different detector sensitivities. 

The result is shown in Fig.~6 in the main Letter. LIGO-LF increases the peak SNR by a factor of 4 relative to aLIGO. While it may still be challenging to detect the effect from a single event, LIGO-LF nonetheless has a promising future in detecting this event via event-stacking. As suggest in Ref.~\cite{Lasky:16}, aLIGO will need $\sim 90$ GW150914-like~\cite{LSC:16} events to be able to achieve a SNR of 5 detection of the memory effect. Accumulating these many events will require aLIGO to operate at full sensitivity for $\sim 10$ years (note that the detection rate calculations in Section~\ref{sec:det_rate} assumes an SNR lower limit of 8 or a range of $z\simeq0.4$ for a $30\,M_\odot$-$30\,M_\odot$ system; restricting to a range of within $z\lesssim 0.1$ will lower the detection rate by a factor of $\simeq 64$). For LIGO-LF, however, only $\sim 25$ events will be sufficient to reach a similar level of detection. As the detection rate of LIGO-LF also increased by a factor of almost 20 relative to aLIGO, it means within a few months of observation with LIGO-LF will be sufficient to achieve a high SNR ($\ge 5$) detection of the memory effect.

\section{Significance of the sub-15\,Hz sensitivity}
Reaching the sensitivity requirement below 15\,Hz is the most technological challenging part of the LIGO-LF upgrade. For example, LIGO-LF requires the masses of the upper suspension stages (TOP, UIM and PUM) to be increased by a factor of 4 or 5 relative to the aLIGO case in order to achieve sufficient attenuation of the seismic motion and the suspension thermal noise (cf. Section~\ref{sec:sus}). This will result in a payload close to the limit of the capacity of the current vacuum chamber piers, and will thus require the suspension system to be carefully centered in the chamber so that the capacity can be fully utilized. Meanwhile, in order to make the angular noise subdominant, the sensitivity of the damping sensors need to be improved by a factor of $\sim 100$. However, currently there are no sensors with noise this low that can be directly integrated to the LIGO suspension system. Lastly, the subtraction factor of the gravity gradient noise is largely uncertain as no direct measurement of this noise is yet available. 

Nevertheless, it is also scientifically rewarding if LIGO-LF can achieve the proposed sensitivity between 5-15\,Hz. In this Section we examine the significance of this band for both astrophysics and detector science. 

\begin{figure}
 \includegraphics[width=0.48\textwidth]{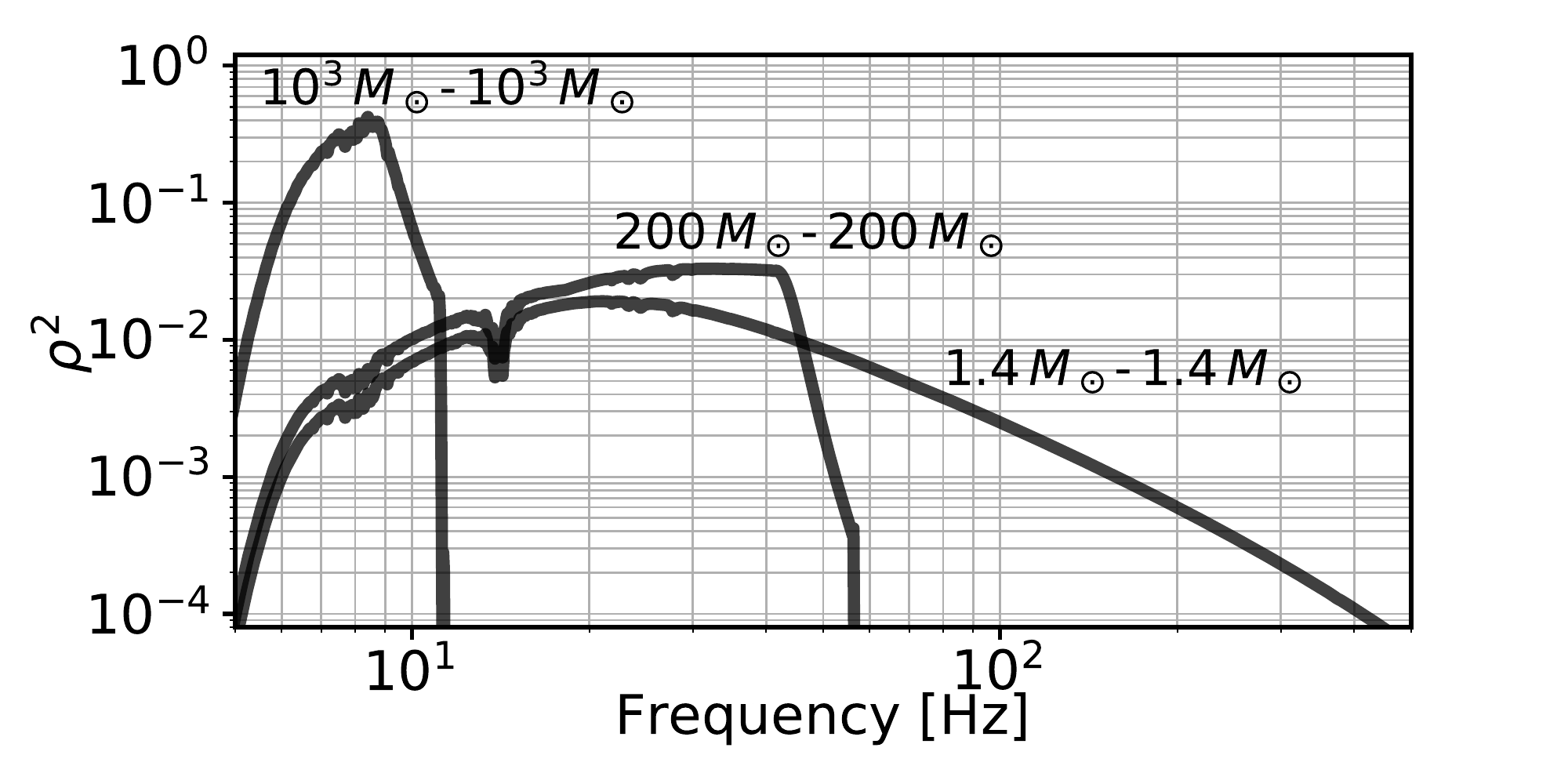}%
 \caption{The square of the SNR density seen in LIGO-LF, $\rho^2(f)$, for binary systems with different total masses. The areas are normalized such that the total integrated SNR for each system is unity, and all the masses refer to the detector-frame masses. Without sensitivity below 15\,Hz, LIGO-LF will not be able to detect systems with total mass above $1000\,M_\odot$, and can hardly measure the inspiral signal for binaries more massive than $400\,M_\odot$. 
 \label{fig:snr_density}}
\end{figure}

In Fig.~\ref{fig:snr_density} we show the square of the SNR density, $\rho^2(f)$, for different binary systems and with the LIGO-LF sensitivity. Here for a piece of GW signal $h(f)$, the square of its SNR density is given by
\begin{equation}
\rho^2(f)=4\frac{h^\ast(f)h(f)}{S(f)},
\end{equation}
where $S(f)$ is the power spectra density of the instrumental noise. For the $10^3\,M_\odot$-$10^3\,M_\odot$ system, its SNR comes entirely from this sub-15\,Hz band. The $200\,M_\odot$-$200\,M_\odot$ system acquires most of its SNR in the 20-40\,Hz band. Nonetheless, as the merger happens roughly at twice the ISCO frequency $f_{\rm ISCO}$, and the $f_{\rm ISCO}=5.5\times(400\,M_\odot/M_{\rm tot})\,{\rm Hz}$, losing the 5-15\,Hz band means that we will lose almost all the signal from the inspiral phase for such systems. Note that the masses here stand for the detector-frame masses. Thus not only IMBHs but also stellar-mass BHs at cosmological distances will be affected. 

\begin{figure}
 \includegraphics[width=0.48\textwidth]{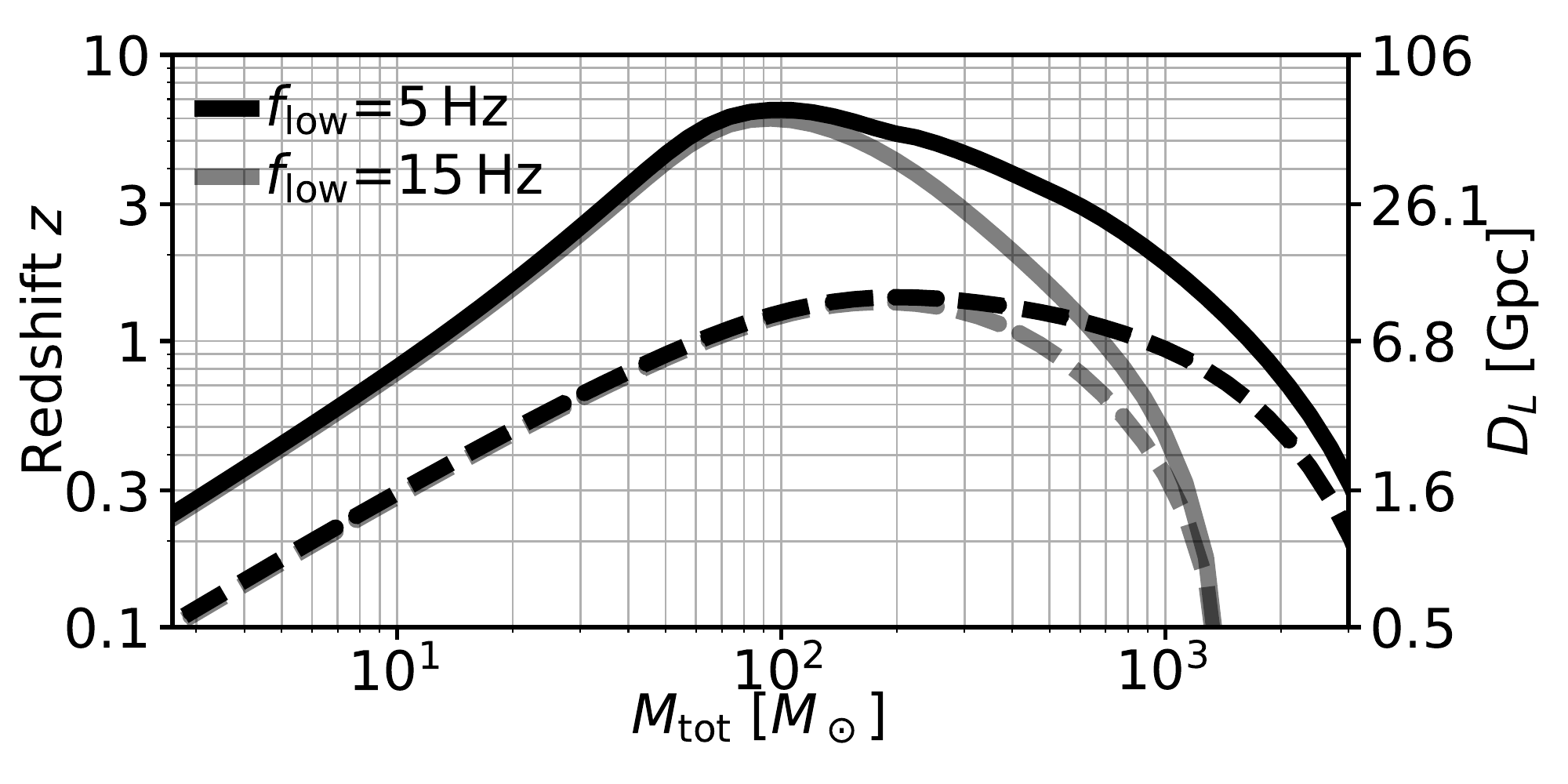}%
 \caption{The detectability horizon (solid) and range (dashed) for binaries with different total mass. In the black curve the full LIGO-LF sensitivity is used, whereas in the grey curve we use only signals above 15\,Hz. Systems more massive than $100\,M_\odot$ are significantly affected by the loss of low-frequency sensitivity.
 \label{fig:z_vs_Mtot_lowF}}
\end{figure}

Similar effects can also be seen in the range-$M_{\rm tot}$ plot as shown in Fig.~\ref{fig:z_vs_Mtot_lowF}. Here the horizontal axis corresponds to mass in the source-frame. As in the main Letter, we assume the sources to have mass ratio of one and zero spin. The range for systems with total mass greater than $100\,M_\odot$ drops significantly if we exclude the sensitivity below 15\,Hz.

The sub-15\,Hz sensitivity may matter even for binary NS systems. While the localization accuracy is insensitive to this band when all of the four HLIV detectors are operating nominally, if during an event we happen to have only HL online, however, then the low-frequency sensitivity becomes crucial.

\begin{figure}
 \includegraphics[width=0.48\textwidth]{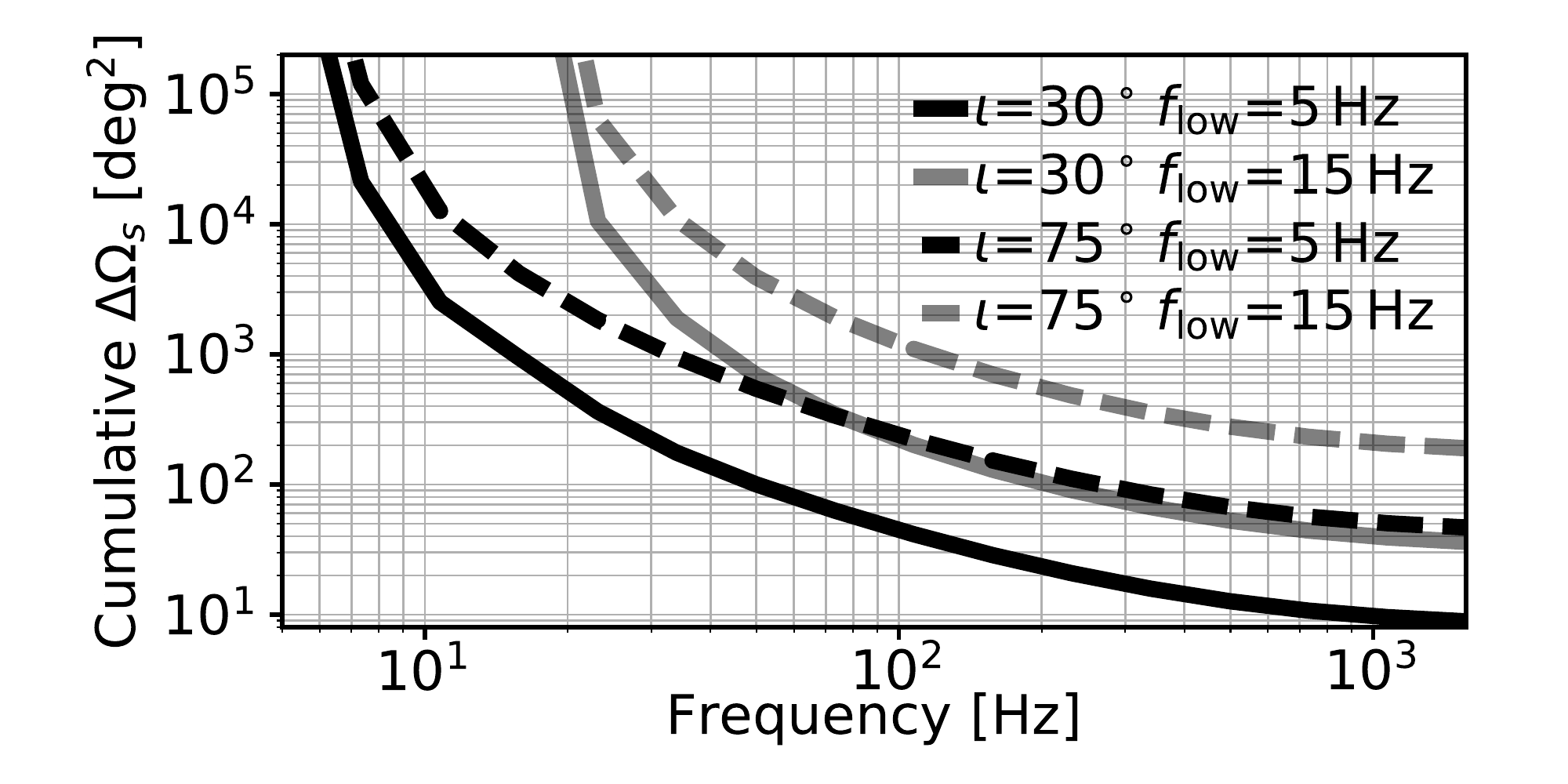}%
 \caption{Localization accuracy for the same sources as we considered in the main Letter and in Section~\ref{sec:bns_loc} of the Supplemental Material. Instead of using the full HLIV network, we consider localizing the source with only two detectors, H and L. In the black curves full LIGO-LF sensitivity is used, while in the grey curves only signals above 15\,Hz are used. Because the 5-15\,Hz band makes it possible to probe the effects due to the Earth's rotation, the source's location can be inferred significantly more accurately with the sub-15\,Hz band than without it. 
 \label{fig:BNS_loc_lowF}}
\end{figure}

In Fig.~\ref{fig:BNS_loc_lowF} we show the cumulative uncertainty area of the same sources as we have considered in the main Letter and in Section~\ref{sec:bns_loc}. When using only the HL network, the localization is significantly more accurate with the full LIGO-LF sensitivity than with only the super-15\,Hz band. This is because with only HL we cannot infer the source location accurately enough from the timing difference between signals measured at the two sites, but have to rely on the modulation of the signal due to the Earth's rotation. As a typical merging NS binary will stay in the 5-15\,Hz band for nearly 2 hours, a significant fraction of a day,
the Earth's rotation thus modifies the antenna responses sufficiently for us to infer the source's sky location.

More importantly, reaching the low-frequency sensitivity of LIGO-LF serves as a critical next step towards the future generation of detectors. As both ET and CE propose to have test masses of at least 200\,kg, the LIGO-LF suspension system naturally serves as a testbed to demonstrate the feasibility of heavy masses and to study the realistic challenges. The inertial seismometers and damping sensors developed for LIGO-LF will likely to be also essential for meeting the ET low-frequency requirement, as ET targets to detect GW at an even lower frequency of 2\,Hz~\cite{Hild:10}. The LIGO-LF will also enable an accurate measurement of the gravity gradient noise which critically determines the lower sensitivity end of CE~\cite{Evans:17}. 

\bibliography{refL}